\documentclass[aps,preprint,preprintnumbers,nofootinbib,showpacs]{revtex4}

\usepackage{amsmath}
\usepackage{amssymb}
\usepackage{mathtools}
\usepackage{dsfont}
\usepackage{graphicx,accents}
\def\bea{\begin{eqnarray}}
\def\eea{\end{eqnarray}}
\def\sea{\nonumber \\&&}
\def\lla{\left\langle}
\def\rra{\right\rangle}

\def\zc{\gamma}
\def\zb{\beta}

\def\ssc{\scriptscriptstyle}
\def\lsim{\mathrel{\raise.3ex\hbox{$<$\kern-.75em\lower1ex\hbox{$\sim$}}} }
\def\gsim{\mathrel{\raise.3ex\hbox{$>$\kern-.75em\lower1ex\hbox{$\sim$}}} }
\begin{document}
%\draft
\preprint{{\vbox{\hbox{NCU-HEP-k093}
%\hbox{Jun 2020}
\hbox{Aug 2021}
\hbox{rev. Oct 2021}
\hbox{ ed. Dec 2021}
}}}
\vspace*{.7in}

%\begin{frontmatter}

\title{\boldmath  Quantum Frames of Reference and the Noncommutative Values of Observables
\vspace*{.2in}}

\author{Otto C. W. Kong}
%\ead{otto@phy.ncu.edu.tw}

\address{ Department of Physics and Center for High Energy and High Field Physics,
%Center for Mathematics and Theoretical Physics, 
National Central University, Chung-li, Taiwan 32054  \\
}

%\cortext[cor1]{Corresponding author.}

\begin{abstract}
\vspace*{.3in}
Based on a recent relational formulation of quantum reference frame
transformations, especially with a case of quantum spatial translations
in particular, we analyzed how the `value' of an observable for a fixed
state change. That is the exact analog of the classical description, for
example, of the value of the $x$-coordinate for a particle decrease by
2 units when we perform a translation of the reference frame putting 
the new origin at $x=2$. The essence of the quantum reference 
frame transformations is to have the quantum fluctuations, and even
entanglement, of the physical object which serves as the (new) reference 
frame, taken into account. We illustrate how the recently introduced 
notion of the noncommutative values of quantum observables gives 
such a definite description successfully. Formulations, and an analysis 
of a case example in qubit systems, of analog transformations for 
observables with a discrete or finite spectrum is also presented. Issues 
about the evolving picture of the symmetry system of all quantum 
reference frame transformations discussed.
\\[.2in]
%\begin{keyword}
\noindent{Keywords :}
Quantum Frames of Reference; Noncommutative Values of Observables;
%Lorentz Covariant Quantum Mechanics; Minkowski Metric Operator; Pseudo-unitary Representation; Pseudo-Hermitian Quantum Mechanics; Symmetry Contraction Limits; Quantum Nonrelativistic and Classical Limits; Quantum Relativity; WWGM Formulation; Coherent State Representation; Noncommutative Spacetime
%Relativity Symmetry, Quantum Relativity, Lie Algebra Contractions, , Quantum Nonrelativistic and Classical Limits
%\PACS 02.20.Qs, 03.65.Ca, 03.65.Fd, 03.65.Ta
%\end{keyword}
%\end{frontmatter}
%\thispagestyle{fancy}
\end{abstract}

\maketitle

\section{Introduction}
The basic idea of relativity in physics is that the values of dynamic 
variables can only be given relative to a frame of reference. Motion
is to be seen as motion with respect to  a frame of reference,
which may be itself in motion when observed from another frame
of reference. Physical laws are to be invariant with respect to the
symmetry of transformations among the choice of admissible class
of frames of references. Naively, a frame of reference is an abstraction 
of an idealized physical system.  In practice, we can only have actual,
less than ideal, physical systems to be used.  That does not raise any
new theoretical concern so long as the latter can be considered
classical. Some implications from the quantum nature of the physical 
system to be used as the frame of reference has been discussed 
back in 1967 \cite{AS}. Before the turn of the century, however, 
the only more notable papers addressing the topic are, apparently,
Refs.\cite{AK,R}. The last one was inspired from consideration of
candidate quantized theory of gravity. The general notion of
so-called relational formulation of physics has been highlighted. 
A simple way to put it is that there is no absolute frame of reference, 
which is really the relativity principle. All physical quantities and
states are abstract notions the explicit description of which is
reference frame dependent. With respect to quantum physics,
described in the usual language, not only the expectation value
of an observable is reference frame dependent, its fluctuations
around the expectation value, or the whole statistics of results 
from projective measurements of an observable for a fixed state, 
would also be. The issue of a quantum frame of references is about 
the latter. The quantum fluctuations of a measuring equipment, 
for example, would give results with fluctuations even when 
measuring what we believe to be a classical object, or what we 
otherwise considered an eigenstate. We do not want to rush to the
conclusion that the notion of a system being quantum or classical
is relative. At least all of us humans seem to be classical enough to 
one another, and most of the macroscopic world looks classical to 
us too. Classical means, here with negligible quantum fluctuations 
and entanglement. It suffices to say that quantum frames of 
reference and transformations  are worthy of serious studies.

Advance in experimental quantum physics in a laser environment
 \cite{B} and otherwise \cite{KBC} challenges measuring from
a physical quantum frame of reference. There are also analyses
on plausible applications of the subject matter, as in Refs.\cite{a1,a2}
for examples. The parallel popularity of theoretical studies on 
the subject matter and related issues is, for example, well 
illustrated by the long lists of references in Refs.\cite{0,1}, which 
are the key background references for our analysis here. We will 
refrain from copying over those long lists. Ref.\cite{0}, following 
Ref.\cite{AK}, focuses much on the conceptually simplest and most 
fundamentally interesting case of reference frames connected by 
a spatial translation in which at least one of the frames is not 
classical. Moreover, it gives a completely relational analysis with 
interesting and important results. Ref.\cite{1}, we think, pushes 
the particular analysis forwards in an important way. A key
inspiration we have taken over from the reference is to have 
a formulation with the reference frame itself taken into account
in the Hilbert space of the relevant states. That allows the
presentation, in the example, of the quantum spatial translation 
as the action of a unitary operator within a single Hilbert space, 
hence exactly as a symmetry transformation,
all in the language of states as in kets and bras. In the classical 
perspective, reference frame transformations are, mathematically,
symmetry transformations as changes of coordinates of the 
physical space or phase spaces of physical systems. The totality
of all admissible reference frame transformations that keeps 
a dynamical theory invariant is the relativity symmetry. The 
latter is considered the most fundamental symmetry, especially
interpreted as one for the physical space or spacetime. The 
notion of quantum reference frame transformations asks for
a modification or generalization of the perspective. We are not 
talking about a speculative new kind of quantum symmetry,
but what is there in the theory and practical applications of 
quantum mechanics awaiting a full understanding.  We still see 
many important issues left to be addressed properly, both 
technically and conceptually. Venturing into the direction is 
our target task here. Other interesting papers with closely 
related studies include Refs.\cite{2,3,4,5}.

The first notable feature of the approach to formulate a quantum
spatial translation, as seeing the position observable of a particle 
$A$ as relative to the position observable of another particle $B$, 
something like $\hat{x}_{\!\ssc A}-\hat{x}_{\!\ssc B}$, is that it 
involves nontrivial action on the momentum observables. 
In fact, it is a canonical transformation that preserves the quantum 
Poisson bracket, effectively the commutator, among observables.
After all, symmetries of a quantum system should act as unitary
transformations on the Hilbert space. Such a transformation also
has a fixed action on all observables, and preserves the Poisson
bracket. A careful analysis of the feature show a logical connection 
with an intuitive perspective of seeing the theory of quantum 
mechanics as one of particle dynamics on a quantum model the 
physical space as the phase space \cite{081} with the position and 
momentum observables as a kind of noncommutative coordinates. 
An important related conceptual notion is the noncommutative 
value of a quantum observable \cite{079} which, among other 
things, admits a rigorous way of seeing such an individual definite 
quantum translation as a generalization of the classical one of 
translating by a fixed value of distance. In the classical case, we  
often consider a simple translation of a coordinate by a fixed 
amount, like $x'=x - a$. With or without explicitly thinking about 
$a$ as the coordinate value of an object as the new frame of 
reference, transcribing the description of a physics phenomenon 
from one using $x$ to one using the new $x'$ as a spatial coordinate 
is effectively a reference frame transformation. We want to look
at the exact analog for the quantum case. When a specific state
of $B$ is known, we want to translate $\hat{x}_{\!\ssc A}$ 
by an `amount' $[\hat{x}_{\!\ssc B}]_\phi$, not as the variable
$\hat{x}_{\!\ssc B}$ but, an explicit `value' specific to the state
as the analog of the real number $a$ of the classical case. That
`value', however, cannot be a single real number. The latter
simply cannot encode the full quantum information about the 
position of $B$ at a fixed state $\left|\phi\rra$ including quantum 
fluctuations and plausible entanglement which are the key interests 
about quantum reference frame transformations. Recall that a
classical reference frame transformation for a quantum system
is to be seen as an approximate description, or idealization, in
which that state $\left|\phi\rra$  of the new reference frame 
$B$ is essentially classical. It is important to note that the latter 
requires not only that fluctuations in  $[\hat{x}_{\!\ssc B}]_\phi$ 
be negligible, but the fluctuations in $[\hat{\mathcal O}_{\!\ssc B}]_\phi$ 
for any observable $\hat{\mathcal O}$ be the same. The
noncommutative value introduced in Ref.\cite{081,079} as an
algebraic representation of the full quantum information a
state contains for an observable is here applied to look at the 
kind of changes in physical quantities, such as $[\hat{x}_{\!\ssc B}]_\phi$, 
under the transformations. Explicit illustrations of how that
encodes changes in the expectation value, the quantum fluctuations, 
and entanglement will be presented.  In short, the main results 
presented in this article is the illustrations, through the various 
examples, of how the change in each noncommutative value answers 
explicitly the question of how the particular physical quantity for 
a specific state changes under a quantum reference transformation.

Let us elaborate more on the notion of the noncommutative
value as a description of the full quantum information involved.
A naive thinking about the full information should, for example, 
encode the full statistical distribution of the corresponding 
projective measurements of the observable for a fixed state. 
That has information about the expectation value as well as the 
quantum fluctuations around it, including the Heisenberg uncertainty. 
That may give a good idea on the quantum amount of translation 
$[\hat{x}_{\!\ssc B}]_\phi$, only which will be able to reveal the 
very interesting quantum features of the transformations studied 
in Refs.\cite{0,1}. Though the latter articles do not explicitly discuss 
such a quantum value translated, the description of the effects of 
the translations as in how specific kinds of states are transformed 
are presented, illustrating the `quantumness' of the change in the
translated position as involving changes in the quantum fluctuations 
and even entanglement. Looking at the individual results on the
changes of the wavefunctions, the quantum value of change is
implicitly there. The notion of a noncommutative value of a 
quantum observable we introduced recently \cite{081,079}, is 
exactly a concrete mathematical way to describe the kind of 
quantum values. In fact, the quantum spatial translation may be 
one of the best place to reveal the nature of the noncommutative 
values as formulated from abstract mathematics. We apply a new 
variant of the formulation of such noncommutative values, which 
best suits the purpose at hand, below to look at the quantum 
spatial translations of Refs.\cite{0,1}, aiming at understanding 
better both the quantum reference frame transformations and 
the noncommutative values.

In the next section, we put down formulations of the quantum spatial 
translations and the key relevant results following from Refs.\cite{0,1}, 
to set the platform for our analysis. Note that our explicit formulation 
has a part, as the representation of the states in the full Hilbert space 
including the old and new reference frames, that is different from that 
of Ref.\cite{1}. We will discuss the merit of the formulation as an 
improvement on the latter.
Sec.\ref{sec3}, is then devoted to presenting and analyzing the 
changes in the noncommutative values of the observables under 
the quantum spatial translations for a couple of illustrative cases. 
Most readers are probably new to the notion of the noncommutative
values. In the beginning of the section, we give an essentially
self-contained presentation of a convenient new variant of it. The
part together with the explicit functional representation under the
use of Schr\"odinger wavefunction presented in the Appendix 
should be enough for following and understanding our analyses.
The fact that the notion be conceptually new, however, means that
one needs to be careful to follow the exact mathematical logic 
involved to avoid misinterpreting results, and may have to bear
with the uncomfortable feeling of dealing with something unfamiliar.
We have to beg the readers patience on that.

After that, we take a detour to look at an analog quantum reference 
frame transformation in a system of qubits in the section to follow. 
This is particularly interesting both for the theoretical and the 
practical consideration. We are interested in the general topic of 
quantum reference frame transformations. The quantum spatial 
translation as one of the conceptually most fundamental and well 
formulated explicitly is really taken as a case example. However, as 
the translation is basically formulated from a picture of translations 
of the set of eigenstates, the fact that the position operator has a 
continuous spectrum distinguishes it from a transformation based
on an observable with a discrete, and especially finite set of
eigenvalues. For a quantum system with a Hilbert space of finite
dimension, that is all we have. The question of the analogous 
transformations is hence of key interest. In the simple case as given
by a system of qubits,  the mathematics involves would be easy,
the corresponding analysis of the noncommutative values and
their changes may have results easier to appreciate and hence 
helpful especially for more skeptical readers to understand the
notion. In a way, our formulation here sketches a basic approach
that can gives any quantum reference frame transformation on
such systems based on the notion of an observable as the `position' 
observable of the object taken to be the new frame of reference in 
the case of the quantum spatial translation. Practically, most of the 
important experimental studies with good precision on quantum 
features of systems have been performed on qubit systems. The 
formulation may then have the theoretical implications checked
experimentally.

In section~\ref{sec5}, we discuss various theoretical issues of 
looking at quantum reference frame transformations as a kind of 
symmetry transformations, especially coordinate transformations.
Comparison with the usual Lie group symmetry
picture is discussed. The key feature of plausible entanglement 
of the system of interest and the old and new frames of reference
as physical system is highlighted as what makes the transformations
different from the Lie group symmetries as in classical reference
frame transformations. However, a noncommutative canonical
coordinate picture of the phase space gives the parallel with the 
classical case and allows a real/complex number coordinate 
description of the quantum transformations. The noncommutative 
values of the changes of observables involved illustrate well that 
we are dealing with a system of symmetries beyond the familiar 
framework. We are only at the beginning of our effort to fully 
understand the mathematical structures involved. Some concluding 
remarks will be presented in the last section. The appendix gives 
the basics on the notion of noncommutative values with the 
Schr\"odinger wavefunction representation of the states, used 
in our analysis, for the first time.

\section{The Spatial Translation of Quantum Frame of Reference}
Ref.\cite{0} approaches the issue completely in terms of observables, 
mostly the position and momentum observables. A spatial translation 
as a change of relative position coordinates as seen from an inertial 
(laboratory) frame $A$ to the relative position coordinates as seen 
from another frame $B$, with a third system $C$ under consideration 
is presented as a canonical transformation first explicitly given as 
\bea&&
\hat{x}_{\!\ssc B}^{\ssc (A)} \longrightarrow  -\hat{x}_{\!\ssc A}^{\ssc (B)} \;,
\qquad
\hat{p}_{\!\ssc B}^{\ssc (A)} \longrightarrow -(\hat{p}_{\ssc A}^{\ssc (B)}+\hat{p}_{\ssc C}^{\ssc (B)}) \;,
\sea
\hat{x}_{\!\ssc C}^{\ssc (A)} \longrightarrow \hat{x}_{\!\ssc C}^{\ssc (B)} - \hat{x}_{\!\ssc A}^{\ssc (B)} \;,
\qquad
\hat{p}_{\!\ssc C}^{\ssc (A)} \longrightarrow \hat{p}_{\ssc C}^{\ssc (B)} \;.
\label{T}
\eea
Note that our notation here is mostly in-line with Ref.\cite{1} instead. 
While $A$, particularly, as a frame of reference for position and 
momentum observables, may have to be a system with some structure 
under any practical consideration, so long as the transformation 
considered is concerned, we only have to address its center of mass 
degrees of freedom, which of course behaves exactly as those for 
a single particle. Similarly for all, we may simply think about $A$, $B$, 
and $C$ each as a quantum particle. The expressions for the position and 
momentum observables each has no components. Generalization to the 
case that each is a three-vector of independent components would be 
straightforward. The transformation as a quantum spatial translation
is easy to appreciate. The part of the position observables read as
classical ones would be exactly what one has in a classical theory.
The part of momentum observables is what is required to make the
full transformation a canonical one, {\em i.e.} to have the Poisson
bracket $\frac{1}{i\hbar} [\cdot,\cdot]$ or all $\hat{x}$-$\hat{p}$
commutators preserved.  Implicitly, the thinking about quantum
reference frame transformations has hidden in it an intuitive 
but formally not so trivial \cite{078} picture of the position and 
momentum observables as (noncommutative) coordinates of the 
phase space for the quantum system. Quantum reference frame
transformations are symmetry transformations of the latter.

A unitary operator
\bea
\hat{S}_x =\hat{\mathcal P}_{\!\ssc AB} e^{i \hat{x}_{\!\ssc B}^{\ssc(A)}\hat{p}_{\!\ssc C}^{\ssc(A)}} \;,
\eea
where $\hat{\mathcal P}_{\!\ssc AB}$ is a parity-swap that sends 
$ \left|x\rra_{\!\ssc B} \otimes \left|y\rra_{\!\ssc C}$ to 
$\left|-x\rra_{\!\ssc A} \otimes \left|y\rra_{\!\ssc C}$,
mapping from 
${\mathcal H}_{\!\ssc B}^{\!\ssc(A)}\otimes {\mathcal H}_{\!\ssc C}^{\!\ssc(A)}$,
the Hilbert space for states of the composite system $BC$
as described from $A$, to 
${\mathcal H}_{\!\ssc A}^{\!\ssc(B)}\otimes {\mathcal H}_{\!\ssc C}^{\!\ssc(B)}$,
the Hilbert space for states of the composite system $AC$
as described from $B$, is given to achieve the above 
operator transformations as 
${\mathcal O} \rightarrow \hat{S}_x {\mathcal O} \hat{S}_x^\dag$
 \cite{0}. Given the fact that 
$[\hat{x}_{\!\ssc I}^{\ssc(A)},\hat{p}_{\!\ssc J}^{\ssc(A)}] = \delta_{\!\ssc I\!J}i$,
($\hbar=1$ units used throughout the paper), we have
$e^{i \hat{x}_{\!\ssc B}^{\ssc(A)} \hat{p}_{\!\ssc C}^{\ssc(A)}}$
naively behaves as a translation in $\hat{x}_{\!\ssc C}^{\ssc(A)}$ 
by the `parameter' $\hat{x}_{\!\ssc B}^{\ssc(A)}$ and as
a translation in $\hat{p}_{\!\ssc B}^{\ssc(A)}$ by the `parameter' 
$-\hat{p}_{\!\ssc C}^{\!\ssc(A)}$ and the subsequent action of 
$\hat{\mathcal P}_{\!\ssc AB}$ finishes the job. 

Note that on 
${\mathcal H}_{\!\ssc B}^{\!\ssc(A)}\otimes {\mathcal H}_{\!\ssc C}^{\!\ssc(A)}$, 
$\hat{x}_{\!\ssc B}^{\ssc(A)}$ is really 
$\hat{x}_{\ssc B}^{\ssc(A)} \!\otimes \hat{I}_{\!\ssc C}$  
and $\hat{p}_{\!\ssc C}^{\!\ssc(A)}$ is really 
$\hat{I}_{\!\ssc B} \!\otimes \hat{p}_{\!\ssc C}^{\ssc(A)}$,
for example. We have, explicitly,
\bea
e^{i \hat{x}_{\!\ssc B}^{\ssc(A)}\!\otimes \hat{I}_{\!\ssc C} .\hat{I}_{\!\ssc B} \!\otimes \hat{p}_{\!\ssc C}^{\ssc(A)}}
  \hat{I}_{\!\ssc B} \!\otimes \hat{x}_{\!\ssc C}^{\ssc(A)}
   e^{-i \hat{x}_{\!\ssc B}^{\ssc(A)}\!\otimes \hat{I}_{\!\ssc C} .\hat{I}_{\!\ssc B} \!\otimes \hat{p}_{\!\ssc C}^{\ssc(A)}}
&=& \hat{I}_{\!\ssc B} \!\otimes \hat{x}_{\!\ssc C}^{\ssc(A)} + \hat{x}_{\!\ssc B}^{\ssc(A)}\!\otimes \hat{I}_{\!\ssc C}
\sea
 \longrightarrow
  \hat{I}_{\!\ssc A} \!\otimes \hat{x}_{\!\ssc C}^{\ssc(B)} - \hat{x}_{\!\ssc A}^{\ssc(B)}\!\otimes \hat{I}_{\!\ssc C}\;,
\eea
where the arrow is the action of $\hat{\mathcal P}_{\!\ssc AB}$.
Similarly, we have
\bea
%e^{\frac{i}{\hbar} \hat{x}_{\!\ssc B}^{\ssc(A)}\!\otimes \hat{I}_{\!\ssc C} .\hat{I}_{\!\ssc B} \!\otimes \hat{p}_{\!\ssc C}^{\ssc(A)}}
e^{i \hat{x}_{\!\ssc B}^{\ssc(A)}\!\otimes  \hat{p}_{\!\ssc C}^{\ssc(A)}}
  \hat{p}_{\ssc B}^{\ssc(A)} \!\otimes \hat{I}_{\!\ssc C}
  e^{- i\hat{x}_{\!\ssc B}^{\ssc(A)}\!\otimes  \hat{p}_{\!\ssc C}^{\ssc(A)}}
 %  e^{-\frac{i}{\hbar} \hat{x}_{\!\ssc B}^{\ssc(A)}\!\otimes \hat{I}_{\!\ssc C} .\hat{I}_{\!\ssc B} \!\otimes \hat{p}_{\!\ssc C}^{\ssc(A)}}
&=& \hat{p}_{\ssc B}^{\ssc(A)} \!\otimes \hat{I}_{\!\ssc C}  - \hat{I}_{\!\ssc B} \!\otimes \hat{p}_{\!\ssc C}^{\ssc(A)} 
\sea
 \longrightarrow
  -\hat{p}_{\ssc A}^{\ssc(B)} \!\otimes \hat{I}_{\!\ssc C} -\hat{I}_{\!\ssc A} \!\otimes \hat{p}_{\!\ssc C}^{\ssc(B)} \;.
\eea
What we have here in terms of the position observable 
of $C$ is a change from its initial position `value' of 
$\hat{I}_{\!\ssc B} \!\otimes \hat{x}_{\!\ssc C}^{\ssc(A)}$, 
as an operator on 
${\mathcal H}_{\!\ssc B}^{\!\ssc(A)}\otimes {\mathcal H}_{\!\ssc C}^{\!\ssc(A)}$,
before the transformation to the final  position `value' of 
$\hat{I}_{\!\ssc A} \!\otimes \hat{x}_{\!\ssc C}^{\ssc(B)}$,
as an operator on 
${\mathcal H}_{\!\ssc A}^{\!\ssc(B)}\otimes {\mathcal H}_{\!\ssc C}^{\!\ssc(B)}$.
We want to think about the change as the difference between 
the final and the initial `values', for which the formulation here
is inadequate even in the abstract as we are dealing with operators 
on different Hilbert spaces. Moreover, the naive difference 
between them as seen in Eq.(\ref{T}) reads 
$\hat{x}_{\!\ssc A}^{\ssc(B)} \otimes \hat{I}_{\!\ssc C}$, or
$-\hat{x}_{\!\ssc B}^{\ssc(A)} \otimes \hat{I}_{\!\ssc C}$, does
not look like having much to do with the position of $C$. These
are puzzles to clarify below.

We give here an alternative formulation of the transformation
in terms of position eigenstates, following but improving on the
work of Ref.\cite{1}. The formulation allows the transformation
to  be seen more directly as a symmetry transformation within the Hilbert space of
${\mathcal H}_{\!\ssc A}\otimes {\mathcal H}_{\!\ssc B}\otimes {\mathcal H}_{\!\ssc C}$,
with the initial and final frames of reference taken as the `states'
$\left|{\bf 0}\rra_{\!\ssc A}$ and $\left|{\bf 0}\rra_{\!\ssc B}$,
{\em i.e.} the zero vectors. We introduce the use of the zero vector 
under the following considerations. A zero vector of course has no 
observable physical properties. Any operator acts on it trivially. That 
corresponds exactly to the idea that a frame of reference does not 
see itself as a dynamical object, hence cannot have a state with any 
nontrivial observable properties. We emphasize that it is no enough
that the description of a composite state of $BC$ as observed from
$A$ has no nontrivial content for its own position, 
$\hat{x}_{\!\ssc A}^{\ssc(A)}$, when the quantum spatial translation 
is concerned. For the case one may think the position eigenstate with 
eigenvalue zero may do. First of all, the idea of writing that state of 
$BC$ as observed from $A$ as a vector within 
${\mathcal H}_{\!\ssc A}\otimes {\mathcal H}_{\!\ssc B}\otimes {\mathcal H}_{\!\ssc C}$
should be a notion independent of which particular quantum
reference frame transformation one may want to formulate. At 
least it is more desirable to have such a consistent formulation. 
In the picture of Eq.(\ref{T}) as presented in Ref.\cite{0}, there is no
$\hat{p}_{\!\ssc A}^{\ssc(A)}$ and $\hat{p}_{\!\ssc B}^{\ssc(B)}$ 
to considered as there is no $\hat{x}_{\!\ssc A}^{\ssc(A)}$ and
$\hat{x}_{\!\ssc B}^{\ssc(B)}$. The zero position eigenstate has
no trivial $\hat{p}_{\!\ssc A}^{\ssc(A)}$ with quantum fluctuations,
and would lead to similar $\hat{p}_{\!\ssc B}^{\ssc(B)}$ after the
quantum translation. One can see that our formulation with the
zero vector is free from that and gives consistent results. Of course 
whatever makes up the physical frame of reference would be 
observed as a usual object from another frame of reference. 

The spatial translation is presented 
as the action of the unitary operator
\bea
\hat{U}_x = \hat{\mathcal S}_{\!\ssc AB}^{\ssc W} \hat{I}_{\!\ssc A} \otimes
    \int\!\!dx'dy' \left|-x'\rra\!\! \lla x'\right|_{\!\ssc B} \otimes \left|y'-x'\rra\!\! \lla y'\right|_{\!\ssc C}  \;,
\eea
which takes a generic state 
\bea
\left|\psi \rra = \left|{\bf 0}\rra_{\!\ssc A} \otimes 
   \int\!\!dxdy  \;\psi (x,y) \left|x\rra_{\!\ssc B} \otimes \left|y\rra_{\!\ssc C}  \;,
\eea
to 
\bea
\hat{U}_x \left|\psi \rra && =    \int\!\!dxdy  \;\psi(x,y) \left|-x\rra_{\!\ssc A} \otimes 
   \left|{\bf 0}\rra_{\!\ssc B} \otimes  \left|y-x\rra_{\!\ssc C}  \;,
\sea
 = \int\!\!dxdy  \;\psi(x,y+x) \left|-x\rra_{\!\ssc A} \otimes 
   \left|{\bf 0}\rra_{\!\ssc B} \otimes  \left|y\rra_{\!\ssc C}  \;.
\label{U}
\eea
%and checks that the latter agrees with $\hat{S}_x \left|\psi \rra$.
$\hat{\mathcal S}_{\!\ssc AB}^{\ssc W}$ is a simple swap sending 
$ \left|z\rra_{\!\ssc A} \otimes  \left|x\rra_{\!\ssc B} \otimes \left|y\rra_{\!\ssc C}$ 
to $\left|x\rra_{\!\ssc A} \otimes  \left| z\rra_{\!\ssc B} \otimes \left|y\rra_{\!\ssc C}$.
It can further be checked explicitly that 
\bea&&
\hat{U}_x \!\!  \int\!\!dz'dx'dy'\; x'  \left|z'\rra \!\! \lla z'\right|_{\!\ssc A}  \otimes  \left|x'\rra \!\! \lla x'\right|_{\!\ssc B}   \otimes \left|y'\rra \!\! \lla y'\right|_{\!\ssc C}  \hat{U}_x^\dag
 = \!\!\int\!\!dzdxdy\; (-x)  \left|x\rra \!\! \lla x\right|_{\!\ssc A}  \otimes \left|z\rra \!\! \lla z\right|_{\!\ssc B}   \otimes \left|y\rra \!\! \lla y\right|_{\!\ssc C} ,
\sea
\hat{U}_x \!\! \int\!\!dz'dx'dy'\; y'  \left|z'\rra \!\! \lla z'\right|_{\!\ssc A}  \otimes  \left|x'\rra \!\! \lla x'\right|_{\!\ssc B}   \otimes \left|y'\rra \!\! \lla y'\right|_{\!\ssc C} \hat{U}_x^\dag
= \!\! \int\!\!dzdxdy\; (y-x)  \left|x\rra \!\! \lla x\right|_{\!\ssc A}  \otimes \left|z\rra \!\! \lla z\right|_{\!\ssc B}   \otimes \left|y\rra \!\! \lla y\right|_{\!\ssc C} ,
\sea
\eea
which are exactly 
$\hat{U}_x \hat{x}_{\!\ssc B}^{\ssc(A)}  \hat{U}_x^\dag = -\hat{x}_{\!\ssc A}^{\ssc(B)}$
and $\hat{U}_x \hat{x}_{\!\ssc C}^{\ssc(A)}  \hat{U}_x^\dag 
    = \hat{x}_{\!\ssc C}^{\ssc(B)} -   \hat{x}_{\!\ssc A}^{\ssc(B)}$.
Now, we can write explicitly 
$\hat{x}_{\!\ssc C}^{\ssc(B)} - \hat{x}_{\!\ssc C}^{\ssc(A)}$ as
$\hat{x}_{\!\ssc C} -  \hat{U}_x \hat{x}_{\!\ssc C}  \hat{U}_x^\dag 
  = \hat{x}_{\!\ssc A} = - \hat{U}_x \hat{x}_{\!\ssc B}  \hat{U}_x^\dag$,
or $\hat{x}_{\!\ssc C}^{\ssc(B)} = \hat{U}_x \hat{x}_{\!\ssc C}  \hat{U}_x^\dag 
   - \hat{U}_x \hat{x}_{\!\ssc B}  \hat{U}_x^\dag$
as the classical analog of $x'_{\!\ssc C} = x_{\!\ssc C} - {x}_{\!\ssc B}$.
Though we focus on the part of the position operators mostly, one can 
also check explicitly for the momentum part. Some detailed calculations 
involving the momentum variables in relation to their noncommutative 
values will actually be presented in the next section. With the
noncommutative values, we can also see below how the 
$\hat{x}_{\!\ssc C}$ value changes by the value of $ - \hat{x}_{\!\ssc B}$,
instead of having only a relation between the operators as the 
dynamical variables.

Anyway, the formulation in terms of unitary transformation 
on a Hilbert space, presented with any basis, of course definitely 
fixed its results on any observable. And the form presented here
is certainly unambiguous and easy to apply to states. Actually,
$\hat{S}_x$ can be written as a unitary transformation on 
${\mathcal H}_{\!\ssc A}\otimes {\mathcal H}_{\!\ssc B}\otimes {\mathcal H}_{\!\ssc C}$
as
\bea
\hat{S}_x = \int\!\!dx'dy'dz' \left|-z'\rra_{\!\ssc B}\!\! \lla z'\right|_{\!\ssc A} \otimes \left|-x'\rra_{\!\ssc A}\!\! \lla x'\right|_{\!\ssc B} \otimes \left|y'-x'\rra\!\! \lla y'\right|_{\!\ssc C}  \;,
\eea
and $\hat{U}_x$ in the same form differs from it as having 
a $\left|z'\rra_{\!\ssc B}$ in the first part, with 
$\hat{S}_x \left|\psi \rra = \hat{U}_x \left|\psi \rra$. The
forms of $\hat{S}_x$ and $\hat{U}_x$ simply present the
transformation without referring to the notion of reference frames. 
Note however that the unitary operators cannot be applied to  
any state involving $\left|{\bf 0}\rra_{\!\ssc B}$, though having
a nontrivial part for $A$ is admissible and may be of interest.

Ref.\cite{0} presents some very illustrative nice pictures of 
the effects of the transformation in its figure~3. We gives here
explicit analytical expressions for the four cases in that figure. 
The results are to be used in our analysis below.  They are, in 
terms of simplified notations, as
\bea&&
\mbox{(a)} :\qquad
\left|x_o\rra  \otimes \int\!\!dy \; \psi(y) \left|y\rra 
 % \sea\hspace*{.8in}
  \qquad\longrightarrow\qquad
   \left|-x_o\rra \otimes  \int\!\!dy \; \psi(y) \left|y-x_o\rra \;;
\sea 
\mbox{(a$^\prime$)} :\qquad\hspace*{.2in}
\left|x_o\rra  \otimes ( c \left|y_{\ssc 1}\rra + s\left|y_{\ssc 2}\rra  ) 
  %\sea\hspace*{.8in}
  \qquad\longrightarrow\qquad
   \left|-x_o\rra \otimes ( c \left|y_{\ssc 1}-x_o\rra + s\left|y_{\ssc 2}-x_o\rra  ) \;;
\sea
%\frac{1}{\sqrt{2}} ( \left|x_{\ssc 1}\rra + e^{i\zeta'}\left|x_{\ssc 2}\rra  ) \otimes  \int\!\!dy \; \psi(y) \left|y\rra 
%   \sea \qquad\longrightarrow\qquad
 %   \frac{1}{\sqrt{2}} \left|-x_{\ssc 1}\rra \otimes  \int\!\!dy \; \psi(y) \left|y-x_{\ssc 1}\rra 
%      + \frac{ e^{i\zeta'}}{\sqrt{2}} \left|-x_{\ssc 1}\rra \otimes  \int\!\!dy \; \psi(y) \left|y-x_{\ssc 2}\rra \;.
\mbox{(b)} : \qquad
\frac{1}{\sqrt{2}} ( \left|x_{\ssc 1}\rra + \left|x_{\ssc 2}\rra  ) \otimes \int\!\!dy \; \psi(y) \left|y\rra 
  \sea\hspace*{.9in}\longrightarrow
\frac{1}{\sqrt{2}} \left( \left|-x_{\ssc 1}\rra  \otimes \int\!\!dy \; \psi(y) \left|y-x_{\ssc 1}\rra + \left|-x_{\ssc 2}\rra  \otimes \int\!\!dy \; \psi(y) \left|y-x_{\ssc 2}\rra \right) \;;
\sea  %\hspace*{.5in}
\mbox{(b$^\prime$)} : \qquad\hspace*{.2in}
\frac{1}{\sqrt{2}} ( \left|x_{\ssc 1}\rra + e^{i\zeta'}\left|x_{\ssc 2}\rra  ) \otimes  ( c \left|y_{\ssc 1}\rra + s\left|y_{\ssc 2}\rra  ) 
  \sea\hspace*{.1in}\longrightarrow
   \frac{c}{\sqrt{2}} \left|-x_{\ssc 1}, y_{\ssc 1}-x_{\ssc 1}\rra 
     + \frac{s}{\sqrt{2}} \left|-x_{\ssc 1}, y_{\ssc 2}-x_{\ssc 1}\rra 
     + \frac{c e^{i\zeta'}}{\sqrt{2}} \left|-x_{\ssc 2}, y_{\ssc 1}-x_{\ssc 2}\rra 
     + \frac{s e^{i\zeta'}}{\sqrt{2}} \left|-x_{\ssc 2}, y_{\ssc 2}-x_{\ssc 2}\rra  \;;
\sea
\mbox{(c)} : \qquad
c \left|x_{\ssc 1}, y_o + x_{\ssc 1}\rra   + s \left|x_{\ssc 2}, y_o + x_{\ssc 2} \rra  
  %\sea\hspace*{.1in}
  \qquad\longrightarrow\qquad
   ( c \left|-x_{\ssc 1}\rra + s\left|-x_{\ssc 2}\rra  )  \otimes \left|y_o \rra \;;
\sea 
\mbox{(d)} : \qquad
 \int\!\!dx  \;\psi (x) \left|x, y_o+x\rra
  \qquad\longrightarrow\qquad
     \int\!\!dx  \;\psi (x)  \left|-x\rra \otimes \left|y_o \rra \;;
\eea
where
${c}\equiv\cos(\frac{\theta}{2}) e^{\frac{-i\zeta}{2}}$ and
${s}\equiv \sin(\frac{\theta}{2}) e^{\frac{i\zeta}{2}}$,
$0 \leq \theta < \pi$, $0 \leq \zeta < 2\pi$, used to write
a generic linear combination of two states. Given the above
presentation, the interpretation of the simplified notations 
should be unambiguous. Case (a) has as the initial state a 
product of position eigenstate for $B$ and a generic state 
for $C$ (together with $\left|{\bf 0}\rra_{\!\ssc A}$). 
The final state maintains being a product state as shown 
(involving $\left|-x\rra_{\!\ssc A}$ and  $\left|{\bf 0}\rra_{\!\ssc B}$).\footnote{
%%% %%%%%%%%%footnote
$C$, $A$, and $B$ in the figure (3 of Ref.\cite{0}) correspond 
to $A$, $B$, and $C$ of our notation, respectively.
}  %%% %%%%%%%%%footnote
We present (a$^\prime$) as a restriction of the generic state 
of $C$ to a linear combination of two position eigenstates, for 
the purposes of matching to results for a system of qubits. 
It  captures the key features of the generic case.  The 
transformation is much like a classical spatial translation 
with the `classical' frame of reference for position represented 
by a position eigenstate when observed from another frame. 
The quantum nature lying in the fact that the position eigenstate
cannot be a momentum eigenstate should be noted. Case 
(b$^\prime$) is a restriction of (b) in exactly the same sense. 
The cases has the transformation of a product state to one with 
nontrivial entanglement (between $A$ and $C$), which may be more
easily appreciated from (b$^\prime$). Initial state for case (c) rather 
generalized somewhat the one in the figure, as a not necessarily 
equal combinations of two perfectly correlated parts of products 
of position eigenstates of $B$, and $C$ with a fixed difference in 
eigenvalue.  The translation to have $B$ as the reference frame 
gives the final state as a product with the part for $C$ as a simple 
eigenstate. The perfect correlation makes all the quantum 
fluctuations of $C$ unobservable from $B$. (d) is really just 
a more general form of (c)  with the same basic feature. Note that 
(b) is much like the inverse of (d). For example, the initial and final
state of (b$^\prime$) with  $c=1$ and  $s=0$ can be identified
essentially with the final and initial state of  (c) with $c=1$ and 
$s=e^{i\zeta'}$, respectively.  We will look at the properties 
of the initial and final states in the transformations in (a) and
(d) in the next section.  

\section{Changing the Quantum/Noncommutative Values of Physical Quantities
\label{sec3}}
Let us first give a representation of the noncommutative value
of a quantum observable $\hat\zb$ on a given physical state.
For the $f_{\!\ssc\hat\zb}(z_n, \bar{z}_n)$ function being the 
expectation value  function of Hermitian operator $\hat\zb$, 
we have
\bea &&
V_{\!\ssc\hat\zb_n} =  \partial_n f_{\!\ssc\hat\zb}
 =   - f_{\!\ssc\hat\zb} \bar{z}_n
   + \sum_m \bar{z}_m (\hat\zb)^m_n\;,
\label{xn}\eea
where $(\hat\zb)^m_{n}$ are the matrix element $\lla m|\hat\zb|n\rra$
over an orthonormal basis $\lla m|n\rra= \delta^m_{n}$, and $z^n$
the complex coordinates of a normalized state
$\left|\phi \rra= \sum_n z^n \left|n\rra$, $n$ runs over the 
dimension of the Hilbert space for the system under consideration. 
The set of $z^n$ also serves as  the homogeneous coordinates of 
the projective Hilbert space as a K\"ahler manifold \cite{BZ}. One 
can check that 
\bea &&
f_{\!\ssc\hat\zb\hat\zc} = f_{\!\ssc\hat\zb}  f_{\!\ssc\hat\zc}   
   + \sum_n V_{\!\ssc\hat\zb_n} V_{\!\ssc\hat\zc_{\bar{n}}} \;,
\sea
 (\hat\zb\hat\zc)^m_{n} =  \sum_l (\hat\zb)^m_{l}  (\hat\zc)^l_{n}  \;,
\sea
V_{\!\ssc\hat\zb\hat\zc_n}  =   - f_{\!\ssc\hat\zb\hat\zc} \bar{z}_n
   + \sum_m \bar{z}_m (\hat\zb\hat\zc)^m_{n}\;,
\label{ncx}
\eea
where $V_{\!\ssc\hat\zc_{\bar{n}}}=\partial_{\bar{n}} f_{\!\ssc\hat\zc}$
is just the complex conjugate of  $V_{\!\ssc\hat\zc_{n}}$
for any (Hermitian) operator $\hat\zc$. We can take the
noncommutative/quantum value $[\hat\zb]_{\phi}$ as
represented by the sequence and complex number values of 
the quantities 
$\{  f_{\!\ssc\hat\zb}, V_{\!\ssc\hat\zb_n},  (\hat\zb)^m_{n} \}$,
evaluated on the state. The noncommutative value of  
an observable as the product $\hat\zb\hat\zc$ is then the 
noncommutative product for two noncommutative values, 
{\em i.e.}  $[\hat\zb\hat\zc]_{\phi}
    = [\hat\zb]_{\phi} \star_{\!\kappa} [\hat\zc]_{\phi}$,
with elements of the sequence as given by the equations above. 
The equation gives the explicit definition of the noncommutative 
product $\star_{\!\kappa}$. For any fixed state, the map from 
the observable algebra to the noncommutative values, taken 
as a noncommutative algebra with the product as given, is 
obviously a homomorphism, maintaining the algebraic relation
among the observables in their values. In particular, for
$\hat\zb=\sum \lambda_m \left|m\rra\!\!\lla m\right|$ at $\left|n\rra$,
we have 
\[  
f_{\!\ssc\hat\zb}= \lambda_n \;,
\quad   
 V_{\!\ssc\hat\zb_m}= \bar{z}_m (\lambda_m - f) = 0\;,
\quad   
(\hat\zb)^m_{l}  = \delta^m_{l}  \lambda_m  \;.
\]
So, an eigenstate of an observable always has all corresponding 
$V_{\!\ssc\hat\zb_n}$ being zero, and degenerate eigenstates for 
an observable have identical noncommutative values. Moreover, 
$\hat\zb=r\hat{I}$ gives $f=r$, have the noncommutative value
behaving essentially as a commutative classical real number 
value. Note that the matrix element $(\hat\zb)^m_n$ can be 
expressed in terms of $f_{\!\ssc\hat\zb}$, $V_{\!\ssc\hat\zb_n}$ and 
${\tilde{k}}_{\!\ssc\hat\zb_{\bar{m}n}} \equiv \partial_{{n}} \partial_{\bar{m}} f_{\!\ssc\hat\zb}$ \cite{079},
hence the full sequence for the noncommutative value can be
obtained from a given expectation value function $f_{\!\ssc\hat\zb}$
on the projective Hilbert space without knowing {\em a priori} 
the explicit operator form of the $\hat\zb$. In fact, one can
check if a function $f(z_n, \bar{z}_n)$ is indeed an $f_{\!\ssc\hat\zb}$
without knowing  $\hat\zb$ \cite{CMP,078}. Moreover, the
classical value $r$ as a constant noncommutative value has also 
${\tilde{k}}_{\!\ssc\hat\zb_{\bar{m}n}}=0$.
 
The particular representation of the noncommutative value, 
which is really a single quantity as an element in a 
noncommutative algebra, is chosen as the optimal one for 
an easy and more transparent illustration of the theoretical 
issue address in this article. The sequence of complex numbers 
representing a $[\hat\zb]_{\phi}$ has three parts. The first is 
the first term which is simply the expectation value. The third 
part is the matrix elements $(\hat\zb)^m_{n}$, which is of 
course independent of the state $\left|\phi \rra$. They can 
be seen here as being there only for the calculation of the 
$\star_{\!\kappa} $-product, mostly the part of 
$V_{\!\ssc\hat\zb\hat\zc_n}$, from Eq.(\ref{ncx}).  The 
 $V_{\!\ssc\hat\zb_n}$ part is the key focus in this article. 
It gives important information about how much the state differs 
from an eigenstate, hence the quantum nature of the quantity 
$[\hat\zb]_{\phi}$. For example, the Heisenberg uncertainty 
characterizing the spread of the eigenvalue results from 
projective measurements about the expectation value is 
given by
\bea
(\Delta\zb)^2_{\!\phi} = f_{\!\ssc\hat\zb^2} - f_{\!\ssc\hat\zb}^2
  = \sum_n |V_{\!\ssc\hat\zb_n}|^2 \;.
\eea

For a convenient analysis of the noncommutative value
for the position operator on states as described by
wavefunction, we give here for the first time the form 
of the noncommutative value in the formulation with 
a Hilbert space in uncountable dimension. The first 
thing is to note that the wavefunction $\phi(x)$ is really 
a collection of infinite number of complex number 
coordinates as $\lla x| \phi \rra$, one for each eigenstate 
$\left| x \rra$ for the value of $x$, $-\infty \le x \le \infty$.  
A complex function can be seen as a collection of complex 
numbers (functional values) one at each point of $x$. 
Then, the matrix elements 
$(\hat\zb)^{x'}_x = \langle x' | \hat\zb | x \rangle$ are 
of course are to be expressed together as a two-variable 
function; for example, $(\hat{x})^{x'}_x = x\delta(x'-x)$.
The coordinate derivatives corresponding to 
$V_{\!\ssc\hat\zb_n}$ may then be expressed together 
as a function which is the functional derivative
$\delta_\phi f_{\!\ssc\hat\zb}$. That is, we have  $[\hat\zb]_{\phi}
    = \{f_{\!\ssc\hat\zb},  \delta_\phi f_{\!\ssc\hat\zb}, (\hat\zb)^{x'}_x\}$
as the noncommutative value. We present in the
appendix a full checking of the consistency of the
picture for the noncommutative value on the observable
algebra. 

A word on the practical physical meaning of the noncommutative
values is in order. Meaning of the noncommutative value for
an observable, or a physical quantity, in physics, like the more
familiar commutative real number value, is supposed to be
logically fully ingrained in its mathematical properties. It is a
definite and state specific algebraic quantity that fully encodes
the information about the observable as obtainable from 
the quantum theory. The commutative real number value does
the same for the classical theory, but fails short in the quantum
theory. In particular, the set of noncommutative values for
all observables of any specific physical state maintains all the
algebraic relations among the observables as variable, 
{\em i.e.} operators. Beyond that, we have for the classical
case the real number value as the value one reads off a
measuring apparatus which experimentalists frequently 
deal with. That is a practical aspect currently missing for the
noncommutative value, but not completely impossible to 
be achieved. In fact, we believe physical quantities are by 
nature noncommutative. The notion of the commutative
real number values serves only as an approximation useful
in the classical setting. The real numbers themselves are 
abstract symbols not to be found in nature. We read off
experimental results from measuring apparatus as real 
numbers only because we have calibrated the output with
a real number scale. And of course there are experimental
outputs showing directly as the plot of a function, for example.
A (real) function is really a collection of infinite numbers of
(real) numbers. The key part of the noncommutative value
description we used here, the $V_{\!\ssc\hat\zb_n}$, is 
much like a function. Actually, the corresponding expression
under the Schr\"odinger wavefunction description of the
state (see the Appendix), which is applied in this section 
to look at the quantum spatial translation 
($V_{\!\ssc\hat\zb_n}(x)  =  \delta_\phi f_{\!\ssc\hat\zb} $), 
is exactly a complex value function. On the whole some 
direct experimental determination of the noncommutative
value of a physical quantity only awaits the ingenuity of
our experimentalists to set up workable schemes for its
achievement. However, without something of the kind, and
that the notion being conceptually theoretically new, most
readers will likely be uncomfortable about its practical
physical meaning. We can only rely on the mathematical
logic presented to speak for itself. 

With the background, we can move on to look at the quantum 
spatial translations of case (a) and (d) in the language of the 
noncommutative values. First thing to note is that an expectation 
value function $f_{\!\ssc\hat\zb}(\phi)$ is of course invariant 
under any unitary transformation. As the terms in the sequence 
representing the corresponding noncommutative value are all 
fixed by the values of the derivatives of $f_{\!\ssc\hat\zb}(\phi)$
for a physical state, the whole noncommutative value should
be invariant. For the quantum spatial translation with
$\hat{x}_{\!\ssc C} \to \hat{x}_{\!\ssc C} - \hat{x}_{\!\ssc A}$,
the operator $\hat{x}_{\!\ssc C}$ before and after the 
transformation are different operators on the same Hilbert
space, as position operator formulated on differently defined
position eigenstate basis which gives the easily appreciable
picture of the translation, as $\left| y \rra \to \left| y -x\rra$.
The noncommutative values of the observable as position of 
$C$ for any physical state changes. That is exactly like the 
translation (of reference frame) in classical physics
${x}_{\!\ssc C} \to {x}_{\!\ssc C} - {x}_{\!\ssc A}$, the explicit
operators describing the same position of $C$ are two 
different (quantum) position coordinate \cite{078} which takes
different values. However, there is a further subtlety as the
$\delta_\phi f_{\!\ssc\hat\zb}(\phi)$ and  $(\hat\zb)^{x'}_x$
terms have values which depend on the choice of basis of the
Hilbert space. We can only compare two noncommutative
values explicitly in the sequence representations when the 
latter have the $\delta_\phi f_{\!\ssc\hat\zb}(\phi)$ and  
$(\hat\zb)^{x'}_x$ terms expressed in the same basis. Say, we 
have to compare the initial and final value of ${x}_{\!\ssc C}$ 
through expressing both noncommutative values in either the 
eigenstate basis before or after the transformation. We illustrate 
in much details for the case of (a), even explicit showing the 
invariance of a noncommutative value under the unitary 
transformation, and some results on the momentum observables.
For case (d) then, we are going to present only the key results.
The two cases can be seen as typical illustrative examples of cases
with or without involving entanglement, respectively. Recall that
(c) is not more than a special case of (d), and the key features of case (b)
correspond well to that of (d) reading in reverse, as we will discuss.

\noindent
$(a):\quad \left|\phi\rra=\left|x_o\rra  \otimes \int\!dy\;  \psi(y) \left|y\rra 
  \quad\longrightarrow\quad
  \left|\phi'\rra =\hat{S}_x   \left|\phi\rra =\left|-x_o\rra \otimes \int\!dy\;    \psi(y) \left|y-x_o\rra $\\
We have an initial state wavefunction $\phi(x,y)= \delta(x-x_o) \psi(y)$. 
For the noncommutative value of $\hat{x}_{\!\ssc B}$ 
on the initial state, we have $[\hat{x}_{\!\ssc B}]_{\phi}^i = 
         \{ x_o, \delta_\phi f_{\hat{x}_{\!\ssc B}}^i , x\delta(x''-x)\delta(y''-y)\}$ 
with
\bea&&
\delta_\phi f_{\hat{x}_{\!\ssc B}}^i 
= -  \bar\phi(x,y) x_o + \int\!dx''dy''\;   \bar\phi(x'',y'') x\delta(x''-x) \delta(y''-y) =0 \; .
\nonumber\eea
This is another explicit illustration of $\delta_\phi f_{\!\ssc\hat\zb}^i  =0$ 
for $\phi$ being an eigenstate of $\hat\zb$.
The noncommutative value of $\hat{x}_{\!\ssc C}$ is
$[\hat{x}_{\!\ssc C}]_{\phi}^i  = 
   \{  y_o, \delta_\phi f_{\hat{x}_{\!\ssc C}}^i , y\delta(x''-x)\delta(y''-y)\}$,
where $y_o$ denote the value of $f_{\hat{x}_{\!\ssc C}}^i$ on $\phi$ and 
\bea&&
\delta_\phi f_{\hat{x}_{\!\ssc C}}^i  = -  \bar\phi(x,y) y_o
    + \int\!dx''dy''\;   \bar\phi(x'',y'') y\delta(x''-x)\delta(y''-y)
\sea\hspace*{.8in}
= (y-  y_o) \delta(x-x_o) \bar\psi(y) \;.
\nonumber\eea
After the transformation, we have the wavefunction 
$\phi'(x',y')= \delta(x'+x_o) \psi(y'+x_o)$ for the final state.  
The new noncommutative values of $\hat{x}_{\!\ssc A}$ 
and $\hat{x}_{\!\ssc C}$ are given by
$[\hat{x}_{\!\ssc A}]_{\phi'}^f  =
     \{ x'_o, \delta_{\phi'} f_{\hat{x}_{\!\ssc A}}^f, x'\delta(x'''-x')\delta(y'''-y')\}$
and  $[\hat{x}_{\!\ssc C}]_{\phi'}^f =
     \{ y'_o, \delta_{\phi'} f_{\hat{x}_{\!\ssc C}}^f, y'\delta(x'''-x')\delta(y'''-y')\}$,
where we have $x'_o= -x_o$, $\delta_{\phi'} f_{\hat{x}_{\!\ssc A}}^f =  0$,
$y'_o= y_o-x_o$, and 
$\delta_{\phi'} f_{\hat{x}_{\!\ssc C}}^f  = (y'-  y'_o) \delta(x'+x_o) \bar\psi(y'+x_o)$.
Hence, we have a noncommutative value of 
$\hat{x}_{\!\ssc C}-\hat{x}_{\!\ssc A}$ as
\[
[\hat{x}_{\!\ssc C}-\hat{x}_{\!\ssc A}]_{\phi'}^f
    =[\hat{x}_{\!\ssc C}]_{\phi'}^f-[\hat{x}_{\!\ssc A}]_{\phi'}^f 
      =\{ y'_o+x_o, %\delta_\phi f_{\hat{x}_{\!\ssc C}}-\delta_{\phi'} f_{\hat{x}_{\!\ssc A}}, 
               (y'-  y_o+x_o) \delta(x'+x_o) \bar\psi(y'+x_o), (y'-x')\delta(x'''-x')\delta(y'''-y')\}\;.
\]
However, the transformation of course gives
$x'=-x$ and $y'=y-x_o$, hence, also $x'''=-x''$ and $y'''=y''-x_o$. 
We have then the confirmation of 
$[\hat{x}_{\!\ssc A}]_{\phi'}^f  
     =\{ -x_o, 0, -x\delta(x''-x)\delta(y''-y)\} = [-\hat{x}_{\!\ssc B}]_{\phi}^i$
and $[\hat{x}_{\!\ssc C}-\hat{x}_{\!\ssc A}]_{\phi'}^f    
            =\{  y_o, (y-  y_o) \delta(x-x_o) \bar\psi(y), y\delta(x''-x)\delta(y''-y)\}
                 =[\hat{x}_{\!\ssc C}]_{\phi}^i $.
The result of $[\hat{x}_{\!\ssc C}]_{\phi'}^f 
  =[\hat{x}_{\!\ssc C}]_{\phi}^i - [\hat{x}_{\!\ssc B}]_{\phi}^i$ 
is the statement that the transformation shifts the
quantum/noncommutative value of the position observable of $C$
by the quantum/noncommutative value of that of $B$. In the case, 
$B$ being an eigenstate, $\delta_\phi f_{\hat{x}_{\!\ssc C}}$ and 
the uncertainty $(\Delta \hat{x}_{\!\ssc C})^2_\phi$ are unchanged.
Actually, without entanglement, the factorization of $\phi(x,y)$ 
and $\phi'(x',y')$ allows the picture of the quantum value of the 
position of $C$  for any initial state of wavefunction $\psi(y)$ 
changes by the fixed noncommutative value of 
$[\hat{x}_{\!\ssc B}]_{\delta(x-x_o)} = [x_o]$, with the 
constant real number $x_o$ taken as a constant 
`noncommutative' value.

Let us also check up changes in the noncommutative values
of the momentum observables. We have $[\hat{p}_{\!\ssc B}]_{\phi}^i = 
         \{ 0, \delta_\phi f_{\hat{p}_{\!\ssc B}}^i , -i\partial_{\!x''}\delta(x''-x)\delta(y''-y)\}$
and $[\hat{p}_{\!\ssc C}]_{\phi}^i = 
         \{ p_o, \delta_\phi f_{\hat{p}_{\!\ssc C}}^i , -i\delta(x''-x)\delta\partial_{\!y''}(y''-y)\}$,
where $p_o$ denote the expectation value of $\hat{p}_{\!\ssc C}$,
$\partial_{\!x''}\delta$ is the derivative of the delta function with respect to the variable, 
$\delta_\phi f_{\hat{p}_{\!\ssc B}}^i =    i  \partial_x\delta(x-x_o) \bar\psi(y)$,
and  
$\delta_\phi f_{\hat{p}_{\!\ssc C}}^i  = (i \partial_y - p_o) \bar\psi(y) \delta(x-x_o)$.
We leave details of the results on the noncommutative values 
of the momentum observable to the appendix. Also given there is 
an explicit presentation of the transformation on $\hat{p}_{\!\ssc B}$,
or the $f_{\hat{p}_{\!\ssc B}}^i$, from which various results for the 
noncommutative values verifying 
$[\hat{p}_{\!\ssc C}]_{\phi'}^f = [\hat{p}_{\!\ssc C}]_{\phi}^i$ and 
$[\hat{p}_{\!\ssc A}]_{\phi'}^f =-[\hat{p}_{\!\ssc B}]_{\phi}^i - [\hat{p}_{\!\ssc C}]_{\phi}^i$.
From the result, we have explicitly $[\hat{p}_{\!\ssc A}]_{\phi'}^f 
        =\{-p_o,\delta_{\phi'} f_{\hat{p}_{\!\ssc A}}^f, , -i\partial_{\!x'''}\delta(x'''-x')\delta(y'''-y')\}$
with 
\[
\delta_{\phi'} f_{\hat{p}_{\!\ssc A}}^f
= -   i  \partial_x\delta(x-x_o) \bar\psi(y) - (i \partial_y - p_o) \bar\psi(y) \delta(x-x_o) \;.
\]
The quantum nature of $B$ as the new reference frame is 
illustrated in the nontrivial $\delta_\phi f_{\hat{p}_{\!\ssc B}}^i$
which contributes to the resulted $\delta_{\phi'} f_{\hat{p}_{\!\ssc A}}^f$
making the latter nontrivial even for the case of trivial 
 $\delta_{\phi'} f_{\hat{p}_{\!\ssc C}}^f$ as in $\psi(y)=e^{ip_oy}$,
{\em i.e.} $C$ in a momentum eigenstate. Moreover, in the case of 
any nontrivial quantum features for the momenta of $B$ and $C$, if we 
have $\delta_\phi f_{\hat{p}_{\!\ssc B}}^i = -\delta_\phi f_{\hat{p}_{\!\ssc C}}^i$,
their contribution to $\delta_{\phi'} f_{\hat{p}_{\!\ssc A}}^f$
cancel one another. The kind of cancellation, however, requires
a perfect correlation of the quantum fluctuations in the two
momenta, which happens only with entanglement in $B$ and $C$
as observed in $A$. The interesting feature is illustrated for the
position observables in case (d) which we are moving onto. 

%\\%%%%
\noindent
$\mbox{(d)} : \quad
 \left|\phi\rra =\int\!\!dx  \;\psi (x) \left|x, y_o+x\rra
  \quad\longrightarrow\quad
       \left|\phi'\rra =\hat{S}_x   \left|\phi\rra =\int\!\!dx  \;\psi (x)  \left|-x\rra \otimes \left|y_o \rra $ \\
%%%%
The initial state wavefunction is $\phi(x,y)=  \psi(x) \delta(y-x-y_o)$. 
We have 
\bea&&
\delta_\phi f_{\hat{x}_{\!\ssc B}}^i =   (x- x_o) \bar\psi(x) \delta(y-x-y_o)  \;,
\sea
\delta_\phi f_{\hat{x}_{\!\ssc C}}^i  =  (y-y_o- x_o) \bar\psi(x) \delta(y-x-y_o) 
  =   (x- x_o) \bar\psi(x) \delta(y-x-y_o)\;,
\nonumber\eea
where $x_o$, again, denotes the value of $f_{\hat{x}_{\!\ssc B}}$,
and the value of $f_{\hat{x}_{\!\ssc C}}$ is then $y_o+x_o$. The 
nature of the results not as products of a function of $x$ and another 
of $y$ is the signature of the nontrivial entanglement here seen 
in the noncommutative values of the observables. In addition, 
the equality of the two is the signature of their perfect correlation.
The final state wavefunction is
$\phi'(x',y')=  \psi(-x') \delta(y'-y_o)$, with
\bea&&
\delta_{\phi'} f_{\hat{x}_{\!\ssc A}}^f =   (x'+x_o) \bar\psi(-x') \delta(y'-y_o)    \;,
%\rightarrow  - (x- x_o) \bar\psi(x) \delta(y-x-y_o) \;,
\sea
\delta_{\phi'} f_{\hat{x}_{\!\ssc C}}^f  =  (y'- y_o) \bar\psi(-x') \delta(y'-y_o) =0\;,
 %  \rightarrow   (y-x- y_o) \bar\psi(x) \delta(y-x-y_o) =0\;.
\nonumber\eea
checking out $[\hat{x}_{\!\ssc A}]_{\phi'}^f   = [-\hat{x}_{\!\ssc B}]_{\phi}^i$
and $[\hat{x}_{\!\ssc C}-\hat{x}_{\!\ssc A}]_{\phi'}^f   =[\hat{x}_{\!\ssc C}]_{\phi}^i $.
The result of 
$[\hat{x}_{\!\ssc C}]_{\phi'}^f  =[\hat{x}_{\!\ssc C}]_{\phi}^i - [\hat{x}_{\!\ssc B}]_{\phi}^i$ 
has zero $\delta_{\phi'} f_{\hat{x}_{\!\ssc C}}^f$ from the cancellation 
$\delta_\phi f_{\hat{x}_{\!\ssc C}}^i  - \delta_\phi f_{\hat{x}_{\!\ssc B}}^i$. 
The perfect correlation between the observables leads to their
difference bearing zero uncertainty, as a result of the cancellation 
of the uncertainties. We can also read the transformation in the
reverse, taking the final product state of $\phi(x',y')$ given in the
reference frame of $B$ as the initial, which would be then expressed 
as the entangled state of $\phi(x,y)$ in the reference frame of $A$ 
upon the quantum spatial translation. The difference between
the $[\hat{x}_{\!\ssc C}]_{\phi'}^f $ and $[\hat{x}_{\!\ssc C}]_{\phi}^i$
above as $- [\hat{x}_{\!\ssc B}]_{\phi}^i=[\hat{x}_{\!\ssc A}]_{\phi'}^f$, 
reads as function of $x'$ and $y'$, as the Hilbert space coordinates
in the position eigenstate basis in the frame of $B$, shows no 
entanglement as $\phi(x',y')$ and $\delta_{\phi} f_{\hat{x}_{\!\ssc B}}^i$
or $\delta_{\phi'} f_{\hat{x}_{\!\ssc A}}^f$ factorize into a product 
of functions of $x'$ and $y'$.  This inverse transformation picture 
essentially illustrates the key features of case (b), {\em i.e.} of
turning a product state into one with entanglement. While the 
difference in $[\hat{x}_{\!\ssc C}]_{\phi'}^f$ and 
$[\hat{x}_{\!\ssc C}]_{\phi}^i$ in case (a) above has factorizable 
expressions in terms of $x$-$y$ or  $x'$-$y'$, here the result has 
a factorizable expression only in terms of $x'$-$y'$. Note that 
though $[\hat{x}_{\!\ssc C}]_{\phi'}^f$  has a nonfactorizable 
expression in terms of $x$-$y$, we cannot say that the latter
show entanglement. $x$ is about the eigenstate or eigenvalue 
of $B$ which has no meaning with $B$ as the reference frame.

\section{Quantum Reference Frame Transformations in Qubit Systems\label{sec4}}
Here in this section, we want to formulate a transformation on
qubit states along the line of the translation by $\hat{x}_{\!\ssc B}$
above and study its results. We take as a transformation by
$\hat\sigma_{\!\ssc 3_{\!B}}$ (as observed from $A$).  Considering 
the analog of Eq.(\ref{U}), we see a nontrivial feature from the
finite dimensional nature of the Hilbert spaces for the individual
parts. For each qubit, we have only two base vectors, $\left|0\rra$ 
and $\left|1\rra$ as eigenstates of $\hat\sigma_{\!\ssc 3}$ with
eigenvalues plus and minus 1. The analog of 
$\left|-x'\rra\!\! \lla x'\right|_{\!\ssc B}$ is clearly 
$\left|0\rra\!\! \lla 1\right|_{\!\ssc B}$ and
$\left|1\rra\!\! \lla 0\right|_{\!\ssc B}$ flipping the sign
of the eigenvalues, hence taking $\hat\sigma_{\!\ssc 3_{\!B}}$
to $-\hat\sigma_{\!\ssc 3_{\!A}}$. However, one cannot put
the eigenvalue shift like $\left|y'-x'\rra\!\! \lla y'\right|_{\!\ssc C}$.
If we start with the state  $\left|00\rra$ (of $BC$), a state for the 
$C$ part with $0$ as the eigenvalue of $\hat\sigma_{\!\ssc 3_{\!C}}$
does not exist in ${\mathcal H}_{\!\ssc C}$. A state with expectation 
value being $0$ then suggests itself. After all, it is about finding
some quantum generalization of classical transformations
and the classical value actually matches better to the
expectation value. For example, neither a position eigenstate
nor a momentum eigenstate should be taken as representing
the classical state among the quantum ones. It is the 
canonical coherent state $\left|x,p\rra$ characterized as
the symmetric minimal uncertainty state labeled by the
expectation values. Under that consideration, the unitary
transformation given by
\bea&&
 \left|{\bf 0}\rra_{\!\ssc A} \otimes \left|00\rra \longrightarrow 
    \left|{\bf 0}\rra_{\!\ssc B} \otimes \frac{1}{\sqrt{2}} ( \left|10\rra + \left|11\rra  ) \;,
\sea
 \left|{\bf 0}\rra_{\!\ssc A} \otimes \left|{\boldmath 01}\rra \longrightarrow 
    \left|{\bf 0}\rra_{\!\ssc B} \otimes \frac{1}{\sqrt{2}} ( \left|10\rra - \left|11\rra  ) \;,
\sea
 \left|{\bf 0}\rra_{\!\ssc A} \otimes \left|10\rra \longrightarrow 
    \left|{\bf 0}\rra_{\!\ssc B} \otimes \frac{1}{\sqrt{2}} ( \left|{\boldmath 01}\rra - \left|{00}\rra  ) \;,
\sea
 \left|{\bf 0}\rra_{\!\ssc A} \otimes \left|11\rra \longrightarrow 
    \left|{\bf 0}\rra_{\!\ssc B} \otimes \frac{1}{\sqrt{2}} ( \left|{\boldmath 01}\rra + \left|{00}\rra  ) \;,
\eea
suggests itself. It gives the following complex coordinate 
transformation on the phase space (from the ones on the
subspace with $\left|{\bf 0}\rra_{\!\ssc A}$ to the ones on 
the subspace with $\left|{\bf 0}\rra_{\!\ssc B}$):
\bea&&
z'{\!\ssc 00}= \frac{1}{\sqrt{2}} (z{\ssc 11}-z{\ssc 10})\;,
\qquad
z'{\!\ssc 01}=\frac{1}{\sqrt{2}} (z{\ssc 11}+z{\ssc 10})\;,
\sea
z'{\!\ssc 10}=\frac{1}{\sqrt{2}} (z{\ssc 00}+z{\ssc 01})\;,
\qquad
z'{\!\ssc 11}=\frac{1}{\sqrt{2}} (z{\ssc 00}-z{\ssc 01})\;,
\eea
as a canonical transformation. On the basic operators,
it gives 
\bea&&
\hat\sigma_{\!\ssc 1_{\!B}}, \hat\sigma_{\!\ssc 2_{\!B}}, \hat\sigma_{\!\ssc 3_{\!B}}
%\hat\sigma_{\!\ssc 3_{\!B}} = \left|{\bf ++}\rra \!\! \lla \mbox{\bf ++}\right| + \left|{\boldmath +-}\rra \!\! \lla {\boldmath +-}\right| 
 %  - \left|-+\rra \!\! \lla -+\right| - \left|--\rra \!\! \lla --\right|
   %\sea\hspace*{.8in}
\longrightarrow 
\hat\sigma_{\!\ssc 2_{\!A}} \hat\sigma_{\!\ssc 2_{\!C}},  \hat\sigma_{\!\ssc 1_{\!A}} \hat\sigma_{\!\ssc 2_{\!C}},
- \hat\sigma_{\!\ssc 3_{\!A}} \;,
  \sea
 \hat\sigma_{\!\ssc 1_{\!C}}, \hat\sigma_{\!\ssc 2_{\!C}},  \hat\sigma_{\!\ssc 3_{\!C}}
%\hat\sigma_{\!\ssc 3_{\!C}}= \left|{\bf ++}\rra \!\! \lla {\bf ++}\right| - \left|{\boldmath +-}\rra \!\! \lla {\boldmath +-}\right| 
  % + \left|-+\rra \!\! \lla -+\right| - \left|--\rra \!\! \lla --\right|
   %\sea\hspace*{.8in}
\longrightarrow
- \hat\sigma_{\!\ssc 3_{\!A}} \hat\sigma_{\!\ssc 3_{\!C}},  
  -  \hat\sigma_{\!\ssc 2_{\!C}},  - \hat\sigma_{\!\ssc 3_{\!A}} \hat\sigma_{\!\ssc 1_{\!C}} \;.
\eea
The transformation of course preserves the commutation
relations among the operators. With the result, the analog
results of quantum reference frame transformations based on
$\hat\sigma_{\!\ssc 1_{\!B}}$ to $-\hat\sigma_{\!\ssc 1_{\!A}}$
and $\hat\sigma_{\!\ssc 2_{\!B}}$ to $-\hat\sigma_{\!\ssc 2_{\!A}}$
are quite obvious. 

With the unitary transformation identified, we go next to look
at the results for a few illustrative cases of initial states parallel
to those analyzed above for the quantum spatial translation.
We can check the two lists versus one another and see how well
their results match. The cases here are labeled exactly as their
matching case from above.
\bea&&
\mbox{(a$^\prime$)} : \qquad
\left|0\rra  \otimes ( c \left|0\rra + s\left|1\rra  ) 
  %\sea\hspace*{.8in}\longrightarrow
   \qquad\longrightarrow\qquad
   \left|1\rra  \otimes \left( \frac{c+s}{\sqrt{2}} \left|0\rra + \frac{c-s}{\sqrt{2}} \left|1\rra  \right) \;,
\sea
\mbox{(b$^\prime$)} : \qquad
\frac{1}{\sqrt{2}} ( \left|0\rra + e^{i\zeta'}\left|1\rra  ) \otimes ( c \left|0\rra + s\left|1\rra  ) 
  \sea\hspace*{.8in}\longrightarrow
 %  \qquad\longrightarrow\qquad
       \frac{c-s}{2} (- e^{i\zeta'}\left|{00}\rra +\left|11\rra ) 
         + \frac{c+s}{2} (e^{i\zeta'}\left|{01}\rra  +  \left|10\rra ) \;,
\sea
%\frac{1}{\sqrt{2}} ( \left|{\bf ++}\rra + e^{i\zeta'} \left|--\rra  ) \\
\mbox{(c)} : \qquad
c \left|{00}\rra + s \left|11\rra  
 % \sea\hspace*{.8in}\longrightarrow   
   \qquad\longrightarrow\qquad
      ( s\left|0\rra + c\left|1\rra  ) \otimes \frac{1}{\sqrt{2}} ( \left|0\rra + \left|1\rra  ) \;.
\nonumber
\eea

Let us check explicitly on the noncommutative values for case (c).
The initial state has  $[\hat\sigma_{\!\ssc 3_{\!B}}]^i$ and 
$[\hat\sigma_{\!\ssc 3_{\!C }}]^i$ as given by
\bea&&
\zb=\hat\sigma_{\!\ssc 3_{\!B}} :
\quad
f_{\!\ssc\hat\zb}=|c|^2-|s|^2 \;, 
\quad
V_{\!\ssc\hat\zb_{n\tilde{n}}}=\{ 2\bar{c}|s|^2, 0, 0, -2\bar{s}|c|^2 \} ,
\quad
(\hat\zb)^{{m}\tilde{m}}_{n\tilde{n}} = \{ 1,1,-1,-1\} ;
\sea
\zb'=\hat\sigma_{\!\ssc 3_{\!C}} :
\quad
f_{\!\ssc\hat\zb'}=|c|^2-|s|^2 \;, 
\quad
V_{\!\ssc\hat\zb'_{n\tilde{n}}}=\{ 2\bar{c}|s|^2, 0, 0, -2\bar{s}|c|^2 \} ,
\quad
(\hat\zb')^{{m}\tilde{m}}_{n\tilde{n}} = \{ 1,-1,1,-1\} .
\nonumber\eea
The identical values of $f_{\!\ssc\hat\zb}=f_{\!\ssc\hat\zb'}$ and 
$V_{\!\ssc\hat\zb_{n\tilde{n}}}=V_{\!\ssc\hat\zb'_{n\tilde{n}}}$ in this case
 is from the perfect correlation between $B$ and $C$. The final state has
\bea&&
\zc=-\hat\sigma_{\!\ssc 3_{\!A}} :
\quad
f'_{\!\ssc\hat\zc}=|c|^2-|s|^2 \;, 
\quad
V'_{\!\ssc\hat\zc_{n'\tilde{n}'}}=\{ -\sqrt{2}\bar{s}|c|^2,-\sqrt{2}\bar{s}|c|^2,\sqrt{2}\bar{c}|s|^2,\sqrt{2}\bar{c}|s|^2\} ,
\sea\hspace*{2.in}
(\hat\zc)^{m'\tilde{m}'}_{n'\tilde{n}'}   = \{ -1, -1, 1, 1 \}  ;
\sea
\zc'=-\hat\sigma_{\!\ssc 3_{\!A}} \hat\sigma_{\!\ssc 1_{\!C}} :
\quad\!
f'_{\!\ssc\hat\zc'}= |c|^2-|s|^2 \;, 
\quad\!
V'_{\!\ssc\hat\zc'_{n'\tilde{n}'}}=\{ -\sqrt{2}\bar{s}|c|^2,-\sqrt{2}\bar{s}|c|^2,\sqrt{2}\bar{c}|s|^2,\sqrt{2}\bar{c}|s|^2\},
\sea\hspace*{2.in}
(\hat\zc')^{m'\tilde{m}'}_{n'\tilde{n}'}
  = \mbox{\tiny$\left(\begin{array}{cccc} 0 & -1 & 0 & 0 \\
    -1 & 0 & 0 &0 \\ 0 & 0 & 0 & 1 \\ 0 & 0 & 1 & 0 \end{array}\right)$}  .
\nonumber
\eea
One can check that
\bea&&
V'_{\!\ssc\hat\zc_{n\tilde{n}}} = 
\left\{  \frac{1}{\sqrt{2}} ( V'_{\!\ssc\hat\zc_{10}}+ V'_{\!\ssc\hat\zc_{11}}),
  \frac{1}{\sqrt{2}} ( V'_{\!\ssc\hat\zc_{10}}- V'_{\!\ssc\hat\zc_{11}}),
   \frac{1}{\sqrt{2}} ( V'_{\!\ssc\hat\zc_{01}}- V'_{\!\ssc\hat\zc_{00}}),
  \frac{1}{\sqrt{2}} ( V'_{\!\ssc\hat\zc_{01}}+ V'_{\!\ssc\hat\zc_{00}}) \right\} \;,
% \sea = \{ 2\bar{c}|s|^2, 0, 0, -2\bar{s}|c|^2 \}
\eea
where the the explicit coordinates are $z'$-coordinates as
$z'_{n'\tilde{n}'}$ for ${n'\tilde{n}'} = 00,01,10,11$ while $n\tilde{n}$
refers to the $z_{n\tilde{n}}$,  ${n\tilde{n}}=00,01,10,11$ as used
throughout the section. The matrix elements of observables in the 
corresponding basis are given directly in the matrix or as a sequence 
of eigenvalues when the latter is diagonal. The result is exactly 
$V_{\!\ssc\hat\zb_{n\tilde{n}}}$, confirming the only not so trivial part 
of $[\hat\sigma_{\!\ssc 3_{\!B}}]^i=[-\hat\sigma_{\!\ssc 3_{\!A}}]^f$.
Similarly, one can easily confirm $V'_{\!\ssc\hat\zc'_{n\tilde{n}}} =V_{\!\ssc\hat\zb'_{n\tilde{n}}}$,
and $[\hat\sigma_{\!\ssc 3_{\!C}}]^i=[-\hat\sigma_{\!\ssc 3_{\!A}}\hat\sigma_{\!\ssc 1_{\!C}}]^f$.
Again, those are nothing more than the confirmation of the
noncommutative value as invariant under a unitary transformation.
$[\hat\sigma_{\!\ssc 3_{\!C}}]^f
   = \{ 0, V'_{\!\ssc{\hat\sigma_{\!\ssc 3_{\!C}}}_{\!n'\tilde{n}'}}, \{1,-1,1,-1\} \}$, 
with $V'_{\!\ssc{\hat\sigma_{\!\ssc 3_{\!C}}}_{\!n'\tilde{n}'}}
    = \{ \frac{\bar{s}}{\sqrt{2}}, -\frac{\bar{s}}{\sqrt{2}}, \frac{\bar{c}}{\sqrt{2}}, -\frac{\bar{c}}{\sqrt{2}} \}$.
The result, especially when rewritten as 
$V'_{\!\ssc{\hat\sigma_{\!\ssc 3_{\!C}}}_{\!n\tilde{n}}}=\{ 0, \bar{c}, -\bar{s},0\}$, 
allows direct comparison versus $[\hat\sigma_{\!\ssc 3_{\!C}}]^{i}$. 
We have, for example, the uncertainties 
$(\Delta \hat\sigma_{\!\ssc 3_{\!C}})^2_i = 2|s|^2|c|^2$ and 
$(\Delta \hat\sigma_{\!\ssc 3_{\!C}})^2_{\!f} = 1$. Product states
like our final state for the case have their coordinates factorizable,
as for example $z'_{n'\tilde{n}'}$ can be written as $z'_{n'}z'_{\tilde{n}'}$
which can be traced to the parallel factorizability of the form
$V'_{\!n'\tilde{n}'}=V'_{\!n'}V'_{\!\tilde{n}'}$.  In our case here, we
have all $V_{\!n\tilde{n}}$ above being not factorizable illustrating
the entanglement of the initial state, while the final state results 
as are $V'_{\!n'\tilde{n}'}$ all clearly factorizable. %Note that the
%expression $V'_{\!\ssc{\hat\sigma_{\!\ssc 3_{\!C}}}_{\!n\tilde{n}}}$
%is not factorizable through $V'_{\!\ssc{\hat\sigma_{\!\ssc 3_{\!C}}}_{\!n'\tilde{n}'}}$.

\section{Transformations of Reference Frames, Coordinates,
and  Symmetries\label{sec5}}  

Let us first recall the theoretical features from reference frame 
transformations in general, and spatial translations in particular,
as we understood in classical physics. The group of admissible
reference frame transformations is a Lie group that is considered
as the relativity symmetry for the theory. The perspective of it
as a symmetry of our physical space-time is used to be considered
fundamental. In fact, the relativity symmetry and the model of
space-time, or spacetime,  are really tied together. For example,
having Lorentz symmetry requires the Minkowski spacetime as
the `correct' model whereas its Galilean limit is to be matched
to the Newtonian space-time. Actually, the Newtonian space-time
and the Minkowski spacetime can be theoretically constructed
each as a coset space of the Galilean and Poincar\'e groups,
respectively (see Ref.\cite{086} for an explicit illustration).
Physics is, however, really about dynamics; and dynamics is
to be considered on states of physical systems, which does not 
include space and time themselves in theories not containing 
gravitational dynamics. In this paper, we focus only on simple 
quantum mechanics, or the quantum version of Newtonian 
mechanics. The space of all physical states for a Newtonian 
particle can also be constructed as a coset space of the Galilean 
group \cite{086}. This particle phase space splits into the 
configuration/position space and the momentum space. The 
single particle configuration space as the space of all possible
positions of the particle is what constitutes a model for the 
physical space.  The phase space is naturally a symplectic 
manifold, which can be seen as a geometric space with symmetry 
transformations naturally to be formulated as Hamiltonian flows
 \cite{086}. In fact, the Poisson bracket structure is essentially
dictated by the Galilean group as a Lie group \cite{ism}. Practical 
dynamics corresponds to the identification of the physical 
Hamiltonian, the energy observable, as among all possible 
generic Hamiltonian functions. Every generator of the 
Galilean group corresponds to an observable whose 
Hamiltonian vector field represents the abstract generator. 
It generates a one-parameter group of Hamiltonian flows on
the phase space preserving the symplectic structure, as well
as an automorphism flow on the observable algebra preserving 
the Poisson bracket \cite{086}. The observable algebra itself
can be seen as the representation of the universal enveloping
algebra,  the group algebra, or some generalization of it, fixed 
by the representations of the generators. Here the observable
algebra represented as functions on the phase space with the 
usual product is commutative. 

The (relativity) symmetry theoretical perspective for a theory
of particle dynamics works even better in the quantum cases 
 \cite{070,087}. For the simple, `nonrelativistic', quantum 
mechanics, starting with an irreducible representation of the
$U(1)$ central extension of the Galilean group, formulation
of all aspects of the dynamical theory is a natural consequence,
which highlights its parallel with the classical theory based on 
the Hamiltonian dynamical picture \cite{CMP,078}. We have
pointed out that the formulation gives an interesting, intuitive,
noncommutative geometrical picture of quantum physics
 \cite{070,081,066,078}, with the comprehensive commutative, 
hence classical, limit retrievable as a contraction of the symmetry
and the representation. All that is only within the familiar symmetry, 
including reference frame, transformation picture based on Lie 
groups.  Not considering a particle with nonzero spin, the irreducible
representation of the quantum relativity symmetry is essentially 
unique. The representation space is just the familiar Hilbert space, 
which is usually called the phase space for the quantum particle. 
The projective Hilbert space as the space of rays in it is the exact 
symplectic manifold each point of which corresponds to a distinct 
physical state. While one can have a classical spatial translation, 
it still has to be formulated as one of the phase space. With the 
$U(1)$ central extension effectively incorporating the Heisenberg
commutation structure, the extended Galilean group simply
does not admit an invariant configuration/position space as 
a representation. Just like the Minkowski spacetime cannot be 
split into Lorentz invariant space and time parts, the quantum
phase space cannot be split into the configuration and the
momentum space parts, though both splitting work under 
the Newtonian approximation. That is the reason for the 
nonexistence of the notion of a quantum configuration space.  
With the perspective that the `physical space' of a theory of 
particle dynamics, or the proper model of it, can only be 
a physically meaning notion as the space of all possible position 
for a free particle, one should conclude that the only proper 
model of the physical space as behind quantum mechanics is 
the phase space \cite{066}. A quantum reference frame is then 
always a reference frame for the latter, and hence a symmetry
transformation as coordinate transformation on the phase
space. The latter as a dual structure to the observable algebra
admits a noncommutative geometric picture with the position
and momentum observables as a basic set of noncommutative
canonical coordinates \cite{081,078} for the phase space ---
exactly as one would intuitively expect. That is the setting
to appreciate why a quantum reference frame transformations
as one we may want to think of as simply such a transformation 
on the physical space like a simple spatial translation has to be 
a canonical transformation and there is no way to look at it 
independent of the phase space. There is no model of physical
space we can think of sensibly with only the position observables
as coordinates. The one with the position and momentum 
observables as coordinates is exactly the right model of the
physical space as behind the theory of quantum mechanics, and
that intuitive quantum phase space agrees the space of physical
states as usually described as the (projective) Hilbert space of
infinite real/complex-number dimension. Each noncommutative
coordinate carries the information of infinite number of 
real/complex-number coordinates, an explicit relation between
which is offer by a representation of each noncommutative
value of noncommutative coordinate as a sequence of infinite
number of complex numbers, like the one used above. 

\begin{table}[b]
%\footnotesize
\begin{center}
\hrule\vspace{.2in}
\begin{tabular}{||c|c|c|c|c|c|c|c|c|c||}    \hline\hline
										& \multicolumn{2}{|c|}{Quantum} 											& Classical		\\ \hline
generators $G$							& \multicolumn{2}{|c|}{$[X_i,P_i]=i\hbar$} 									&    $[X_i,P_i]=0$ \\
 \hline
coordinates 								& $\hat{x}_i$, $\hat{p}_i$ 			& $\Re(\psi(x_i))$, $\Im(\psi(x_i))$ 		& $x_i$, $p_i$\\  %\hline
observables 							& $\hat\zb=\zb(\hat{x}_i,\hat{p}_i)$ 	& {$f_{\!\ssc\hat\zb} (\psi)$} 			& $f(x_i,p_i)$ \\
$U_{\!s}=e^{\frac{s}{i\hbar} G_{\!s}}$		& $G_{\!x'_i} = -\hat{p}_i$, $G_{\!p'_i} = \hat{x}_i$		
					& $d \left|\psi \rra= {\frac{ds}{i\hbar}{G}_{\!s}}  \left|\psi \rra$									& 
\\
										& $U_{\!x'_i}^\dag \hat{x}_i U_{\!x'_i} = \hat{x}_i + x'_i$						
					& 	 	$G_{\!x'_i} =i\partial_{\!x_i}$, $G_{\!p'_i} = {x}_i$											&   $U_{\!x'_i}(x_i,p_i) =(x_i+x'_i,p_i)$ 
\\
										& $U_{\!p'_i}^\dag \hat{p}_i U_{\!p'_i} = \hat{p}_i + p'_i$		
					&		$\frac{d\psi(x_i)}{ds} = \frac{1}{i\hbar} G_{\!s}\psi(x_i)$									&  $U_{\!p'_i}(x_i,p_i) =(x_i,p_i+p'_i)$ 
\\
$-\frac{1}{i\hbar}\tilde{G}_{\!s}$ 			& $\mathcal{X}_{\!\ssc\zb}= \frac{1}{i\hbar}[\cdot, \hat\zb]\equiv\frac{1}{i\hbar}[\cdot, G_{\!s}]$ 
					& 		$X_{\!\ssc\hat\zb}  = \frac{1}{i\hbar}\int \!dx\; [ \bar{V}_{\!\ssc\hat\zb}(x)\delta_{\ssc\bar\psi} - {V}_{\!\ssc\hat\zb}(x)\delta_{\ssc\psi}]$	%= \{ \cdot, f_{\!\ssc\hat\zb} \}									
															&  $X_{\!\ssc f} = \{ \cdot, f \}$ 
\\
					& 	$\mathcal{X}_{\!\ssc\hat{x}_i} = -\partial_{\!\ssc\hat{p}_i}$,  $\mathcal{X}_{\!\ssc\hat{p}_i} = \partial_{\!\ssc\hat{x}_i}$
%										& $X_{\!\ssc {f_{\!\hat{x}_i}}}=-i(x_i-x_o)	\psi(x_i) \delta_{\!\ssc\psi}$		
										& $X_{\!\ssc\hat{x}_i} (f_{\!\ssc\hat{p}_j})=-i\hbar\delta_{ij} = -X_{\!\ssc\hat{p}_j} (f_{\!\ssc\hat{x}_i})$,  		
															&	${X}_{\!\ssc{x}_i} = -\partial_{\!\ssc{p}_i}$,  ${X}_{\!\ssc{p}_i} = \partial_{\!\ssc{x}_i}$ 
%\\
%									&	& $X_{\!\ssc {f_{\!\hat{p}_i}}}=-i[-i\partial_{x_i}-p_o]\psi(x_i) \delta_{\!\ssc\psi}$ &
\\
				& 	$d\hat\zb' = -\frac{1}{i\hbar}\tilde{G}_{\!s}(\hat\zb') ds$ 	& $df_{\!\ssc\hat\zb'} = -\frac{1}{i\hbar}\tilde{G}_{\!s}(f_{\!\ssc\hat\zb'}) ds$
															& $df' =-\frac{1}{i\hbar}\tilde{G}_{\!s}(f') ds$
\\
\hline\hline
\end{tabular}
\caption{\footnotesize
Lie group theoretical picture of symmetries in quantum and classical 
dynamics \cite{070,078,086}: The symmetry action of each infinitesimal 
generator and the one-parameter group on the phase space and observable 
algebra, with observables as functions on the phase space coordinate variables,
is given. We show only, for explicit examples of abstract generators, $X_i$ and $P_i$. 
Other observables as symmetry generators follow the basic pattern. The physical 
dynamics in particular is given the Hamiltonian as the energy observable and
the corresponding real parameter $s$ is the Newtonian time. Classical results 
are exact symmetry contraction (naively, $\hbar\to 0$) limits of the
quantum results, directly for the observable algebra.
%$e^{-\frac{s}{i\hbar}\tilde{G}_{\!s}} (\hat\zb') =  e^{s\mathcal{X}_{\!\ssc\zb}} (\hat\zb')
%= e^{-\frac{s}{i\hbar}\hat\zb}   (\hat\zb') e^{\frac{s}{i\hbar}\hat\zb}  $ ;
}
\end{center}
\end{table}

We summarize the usual Lie group theoretical symmetry picture 
for quantum and classical dynamics in Table~1, with the quantum
case illustrated in matching correspondence of the noncommutative 
coordinate and (commutative) complex number coordinate results.  
Each one-parameter group of unitary transformation generated by
a Hermitian operator on the Hilbert space is a Hamiltonian flow
and the preserved Poisson bracket among functions on the 
latter is the exact correspondence of the operator Poisson 
bracket given by $\frac{1}{i(\hbar)}$ times the commutator.
Each observable $\hat\zb$ corresponds to the expectation 
value function $f_{\!\ssc\hat\zb}$ collection of which, with 
the noncommutative product $\star_{\!\ssc\kappa}$ is an
exact copy of the observable algebra itself. However, a
generic Hamiltonian function of $f(z_n,\bar{z}_n)$ or $f(\psi)$
can be matched to an observable as an operator only if the 
flow generated preserves the full Ka\"hler structure \cite{CMP}.  
Natural sets of complex coordinates are obtained as the set of 
expansion coefficients in terms of sets of orthonormal basis for
the Hilbert space. Such a set of complex coordinates are canonical 
in the sense that the real and imaginary parts give exactly pairs 
of canonical real coordinates. Some of the notation and results 
may be easier to appreciate through checking Appendix. 
The Lie group symmetries can be seen, as illustrated, in the 
noncommutative geometric point of view with the position 
and momentum observables taken as phase space coordinates. 
The phase space and the observable algebra as functions on it can 
be obtained from an irreducible representation of the relativity 
symmetry in the quantum case. A general symmetry generator, 
however, can be a generic Hamiltonian function as an observable, 
outside the Lie algebra of the relativity symmetry. The 
noncommutative geometric point of view provides also 
the natural setting to look at the quantum reference frame 
transformations we are interested in here. An important 
point to note though is that, unlike the Lie group symmetries, 
the quantum reference frame transformations for a system 
cannot be considered on its phase space alone. As we have seen 
above, the plausibility of nontrivial entanglement between 
the system and the old or new frame of reference as another 
physical system demands a more complicated formulation. 
It is still interesting to compare the new kind of symmetry 
transformations with the old ones.

From the above, we see that reference frame transformations
in quantum mechanics are symmetries of the phase space. For
the Lie group theoretical ones as presented in Table~1. We have,
including the classical spatial or momentum translations, being
given in terms of one-parameter groups of Hamiltonian flows
which correspond to automorphism flows on the observable
algebra preserving the Poisson bracket. A `classical' spatial
translation as given by the unitary operator $e^{ia\hat{p}}$
for a fixed real number $a$ is indeed, like its correspondence
in classical mechanics, a canonical transformation. However,
it should be emphasized that the real variable $x$ is {\em not
a spatial coordinate} for quantum mechanics. A spatial coordinate 
should be an observable and a coordinate to the phase space as
a representation of the relativity symmetry. The true spatial 
coordinate is exact the quantum counterpart of the classical $x$,
{\em i.e.} $\hat{x}$. The  `classical' spatial translation takes
$\hat{x}$ to $\hat{x}-a$, with the noncommutative value for
each state having only the first term, as the expectation value,
in the sequence representation shifted by $a$, hence the 
quantum fluctuations are unchanged. The action on the 
wavefunction $\psi(x)$ is the action on the latter as a set of 
complex number coordinates of the phase space. The quantum 
spatial translation considered above shifts the noncommutative 
value $[\hat{x}_{\!\ssc C}]_{\ssc\phi}$ for each `state of particle 
$C$' by a noncommutative value which is taken as the
$[\hat{x}_{\!\ssc B}]_{\ssc\phi}$ with $B$ as the new reference
frame. However, the state involved is really a state of at least 
the composite system of $B$ and $C$. With entanglement, not 
only that we cannot talk about a state of $C$ in itself, we cannot 
even talk about a generic state of the system with a fixed 
$[\hat{x}_{\!\ssc B}]_{\ssc\phi}$. The transformation hence can 
only be given as a single unitary operator without any parameter 
dependence or a notion of generator. Combining such 
transformations is not a problem. It is straightforward to 
check that two successive quantum spatial reference frame 
transformations $A \to B \to D$ gives exactly the transformation 
$A \to D$ for example, though one has to enlarge the composite 
system to consider more and more parts \cite{AK,0,1}. We have 
seen also that in the special case with no entanglement in both 
the initial and final states involved, such a description is admissible. 
But it looks like the only such case has $B$ in a position 
eigenstate. The case taken with the eigenvalue as a real variable 
parameter matches exactly to the `classical' spatial translation 
when effect of the transformation on the new nontrivial state 
description of the old reference frame as a system is neglected. 

Going to the limit when the quantum reference frame considered
becomes classical has been well analyzed, for example in 
Ref.\cite{LMB}. The latter article analyzes the (reference frame
transformation) symmetry picture in quantum physics rigorously
though only in terms of applying the symmetry to the physical
system, instead of formulating description of the system and the
theory about it from the language of representation theory as
mentioned above. It tackles in detail how practical considerations
of the symmetry transformations as applied between physical
frames of reference, with physical quantities defined accordingly
as relative to the frames, may realize the mathematical idealization 
in which no reference frame is explicitly considered. The relevant
symmetry is naturally taken essentially as one of a Lie group. Of
course the results are qualitatively as one would naively expect. 
When the reference frame can be well approximated as a classical
one, the idealized mathematical symmetry picture of the Lie group, 
as described above, is well 
realized. The symmetry transformation picture analyzed there
is not one of our changes in the noncommutative values but the
more conventional one of changes in the statistical distributions
of the projective measurement results. The noncommutative
values for the observables obviously reduce to the classical 
real number values when the physical state descriptions of the
reference frames involved can be taken as essentially classical.
Moreover, it is interesting to note that the key mathematical
objects involved for the physical properties of an observable,
as on a state, in the analysis of Ref.\cite{LMB} is really the 
expectation value function, from which the corresponding
statistical distribution result is to be retrieved. Since the
noncommutative value is really like a local representation
of the expectation value function, our noncommutative values 
of changes picture is rather fully compatible with the analysis. 

It should be emphasized that the canonical transformations as 
described in terms of the position and momentum operators
 \cite{AK,0,1} are also special cases of real/complex number
canonical transformations of the (projective) Hilbert space as 
a symplectic/K\"ahler manifold \cite{078}.  They are special in 
the sense that they are taking the noncommutative canonical 
coordinates to noncommutative canonical coordinates with
truly nontrivial changes in the noncommutative values. Of 
course here we are talking about the phase space of the 
composite system. Note
that a noncommutative value is really an element of the 
noncommutative algebra of such values, and as such is a fixed 
abstract quantity as kind of a noncommutative analog of a real 
number. However, any particular representation of it as a 
sequence of complex numbers, we can see, is dependent on 
the choice of basis of the Hilbert space. Especially in relation 
to the quantum reference frame transformations discussed here, 
we have issues about choice of coordinates at different levels. 
The transformations can be seen as transformations of the set
of noncommutative coordinates, hence changes of their 
noncommutative values. The set of noncommutative coordinates 
is also a set of basic observables to which any observable in the 
observable algebra is like a function of. The transformations 
hence change the representation of all observables, and their 
noncommutative values. The basis of the Hilbert space defined
using a set of eigenstates then changes along with it. To 
compare two noncommutative values, we have to use sequence 
representations based on the exact same basis, as done above.

\section{Concluding Remarks}
We have presented above a picture of quantum reference frame
transformation, mostly illustrated through the example of a quantum 
spatial translation, focusing on its effects on the particular properties
of individual states transformed, under our improved formulation of
the transformation. When the transformation takes a description of 
the position observable $\hat{x}_{\!\ssc C}$ of $C$ from one given 
under reference frame $A$, as $\hat{x}_{\!\ssc C}^{\ssc (A)}$,
to one given under reference frame $B$, as
$\hat{x}_{\!\ssc C}^{\ssc (B)}-\hat{x}_{\!\ssc A}^{\ssc (B)}$,
we give an explicit way to describe that relative position and the
relative changes in the `value' of $\hat{x}_{\!\ssc C}$ as the exact
parallel of the familiar solid cases in classical physics. In the 
classical case, in the transformation 
${x}_{\!\ssc C}^{\ssc (A)} \to {x}_{\!\ssc C}^{\ssc (B)}-{x}_{\!\ssc A}^{\ssc (B)}$,
the value of ${x}_{\!\ssc C}$ changes by an amount given by
${x}_{\!\ssc C}^{\ssc (B)}-{x}_{\!\ssc C}^{\ssc (A)} = {x}_{\!\ssc A}^{\ssc (B)}
= -{x}_{\!\ssc B}^{\ssc (A)}$, which is not just the mathematical
relation between those observables as variables. When 
${x}_{\!\ssc A}^{\ssc (B)}$ is known to have a value, say $2$,
we have ${x}_{\!\ssc C}^{\ssc (B)}={x}_{\!\ssc C}^{\ssc (A)}+2$
which gives explicit results like ${x}_{\!\ssc C}^{\ssc (B)}=5$ for
${x}_{\!\ssc C}^{\ssc (A)}=3$.  To answer that question of how the
particular physical quantity for a specific state changes under the 
quantum reference transformation, we illustrate that the 
noncommutative values of (quantum) observables serves the 
purpose well. In the case of (d) for example, analyzed in 
section~\ref{sec3}, the initial value of $[\hat{x}_{\!\ssc C}]_{\phi}^i$ 
(for $\hat{x}_{\!\ssc C}^{\ssc (A)}$) changes to the final value of
$[\hat{x}_{\!\ssc C}]_{\phi'}^f$  (for $\hat{x}_{\!\ssc C}^{\ssc (B)}$)
with the change (as $ [-\hat{x}_{\!\ssc B}]_{\phi}^i$) in the case
can be seen given by the real number expression $y_{o}-(y_{o}+x_{o})=-x_{o}$
and the functional expression 
$0-(x- x_o) \bar\psi(x) \delta(y-x-y_o)= -(x- x_o) \bar\psi(x) \delta(y-x-y_o)$.
The first one is the relation between the expectation values
which changed by an amount $-x_{o}$. The functional expression 
has an initial nontrivial $V_{\hat{x}_{\!\ssc C}}(x,y)$, as 
$\delta_\phi f_{\hat{x}_{\!\ssc C}}^i$, completely removed in the
corresponding final expression, as $\delta_{\phi'} f_{\hat{x}_{\!\ssc C}}^f$,
which is exactly the zero function due to the cancellation of the
identical $V_{\hat{x}_{\!\ssc B}}(x,y)$ part. One sees no fluctuation
in (any classical description of) the position of $C$, nor any
entanglement feature. The initial state of the case is entangled 
with an exact correlation between ${\hat{x}_{\!\ssc C}}$ and 
${\hat{x}_{\!\ssc B}}$ but far from any position eigenstate. Under 
the, not quite correct, classical geometric language, $B$ and
$C$ are always separated by the Newtonian distance $y_{o}$.
Observing ${\hat{x}_{\!\ssc C}}$ using $B$ as the reference frame
hence, intuitively and otherwise, gives $y_{o}$ as the complete
answer. The initial state description (as observed from $A$) is 
transformed into the final state described as the ${\hat{x}_{\!\ssc C}}$
eigenstate. Entanglement between
$C$ and either the initial frame of reference $A$ or the final
frame of reference $B$ is more the rule than the exception,
a noncommutative value of $\hat{x}_{\!\ssc B}^{\ssc (A)}$, for
example, may not be fixed without knowing the composite state
$BC$ as described from $A$. But with the latter information, we 
do have a definite noncommutative value for $\hat{x}_{\!\ssc B}^{\ssc (A)}$.
In the case there is no entanglement between $B$ and $C$,
a definite noncommutative value for $\hat{x}_{\!\ssc B}^{\ssc (A)}$,
or any observables of $B$, can be given independent of the
state of $C$ as well as in terms of the product state of $BC$.

The information about quantum fluctuations and entanglement
concerning any particular physical quantity is fully encoded in the
mathematical description of the state. The information is quantum
in nature and cannot be fully represented or modeled by a single
real number value as in classical physics. The noncommutative
value, as an element of a noncommutative algebra, however,
encodes that full information.  The explicit results on the 
transformations illustrated that clearly. At least from the theoretical
point of view, the noncommutative value of an observable for
a fixed state described under a choice of frame of reference and 
system of coordinates of the phase space, is completely and
definitely determined. When a number of observables satisfying
a certain algebraic relations as dynamical variables, that exact 
relation is preserved among their noncommutative values for
any particular fixed state, like the (real number) values of the
classical observables. Under a quantum reference frame
transformation, the state description is changed. The mathematical
description of any particular physical quantity of the state has hence
to be changed. Unlike a classical reference frame transformation 
which only changes the expectation value of the observable, the
quantum transformation may change every aspect of a state.  Such 
a change as given by the corresponding noncommutative value
again encodes the full information involved, from which one can 
also read off the changes in the quantum fluctuations and  
entanglement. Without the use of the notion of the noncommutative
values, changes in the description of any particular physical properties 
under a quantum reference frame transformation are only given as 
state independent observable relations \cite{0,1}. Tracing how some
states change under the transformation, as for example presented
in figure ~3 of Ref.\cite{0} can illustrate the very interesting features.
The noncommutative values calculations answer all those explicitly.
The picture also helps to demonstrate that our newly introduced
notion of the noncommutative values at least serves as a very useful
model for the `value' of the quantum observables as an exact, hence
mathematically abstract, description of the corresponding physical
properties for specific quantum states. 

We have discussed how a quantum reference frame
transformation is to be seen as a symmetry transformation or 
coordinate transformation on the quantum phase space. From 
the perspective of the more familiar Lie group theoretical 
formulation of relativity symmetry, the Hilbert space picture 
of the quantum phase space is essentially the only irreducible
representation for a spin zero particle. The notion of the
configuration space for a single particle system as the space
of all possible positions of the particle, hence the physical 
space, does not exist more than as a part of the phase space.
The phase space has to be taken as the geometric model for
the physical space. The position and momentum observables
$\hat{x}_i$ and  $\hat{p}_i$ can be seen, as intuitively expected,
to be a kind of coordinates for the phase space. A quantum
reference frame transformation is a transformation of such
system of noncommutative coordinates hence a symmetry  
of the phase space. Exactly as in the classical case, they are 
canonical transformations preserving the corresponding
Poisson bracket, though here have to be formulated on the
phase space of the composite system with the reference frame(s).
The phase space symmetries are unitary transformations of
the Hilbert space. However, the quantum reference frame 
transformation cannot be formulated as individual ones, for
example one for each translation of $\hat{x}_{\!\ssc C}$ by
a fixed `value' of $\hat{x}_{\!\ssc B}$ unless there is no
entanglement between $B$ and $C$. Even in the latter case,
entanglement of the transformed state of $C$ with the old
reference frame $A$ as a quantum system seen from $B$
is often involved. So, we have general to deal with the
quantum translation of 
${x}_{\!\ssc C}^{\ssc (A)} \to {x}_{\!\ssc C}^{\ssc (B)}-{x}_{\!\ssc A}^{\ssc (B)}$
for any state of the composite system as a single 
symmetry transformation. 

While composing two quantum reference frame transformations
like ${x}_{\!\ssc C}^{\ssc (A)} \to {x}_{\!\ssc C}^{\ssc (B)}-{x}_{\!\ssc A}^{\ssc (B)}$
followed by ${x}_{\!\ssc C}^{\ssc (B)} \to {x}_{\!\ssc C}^{\ssc (D)}-{x}_{\!\ssc B}^{\ssc (D)}$
is straightforward, what is more interesting to combine 
quantum reference frame transformations of different kinds,
like a quantum spatial translation and a quantum momentum
translation, or even a kind of quantum rotations. One can easily
appreciate how a quantum momentum translation can be 
formulated in a parallel manner, and goes further to the case
of three independent pairs of $\hat{x}_i$ and  $\hat{p}_i$.
The angular momentum observables can be used to formulate
the quantum rotations. Our presentation of quantum reference 
frame transformations for qubit systems  serves as a good 
illustration on how to proceed in the case of based on 
observables with a discrete or finite list of eigenvalues. We 
believe we have essentially given the framework to formulate
a quantum reference frame transformation from an initial frame
$A$ to a new frame $B$ based on  any observable $\mathcal{O}$, 
{\em i.e.} under  $\mathcal{O}_{\!\ssc A}^{\ssc (B)}=-\mathcal{O}_{\!\ssc B}^{\ssc (A)}$
and the proper matching transform for  $\mathcal{O}_{\!\ssc C}$. 
We are hence at the starting point of looking into 
the system of symmetries of the totality of all quantum reference 
frame transformations of a given physical system. Further studies
in the direction, together with the examinations of the results in
terms of the noncommutative values for particular states, is 
a task on the table.

On the more technical side, besides the key focus of the analyses of the 
changes in the noncommutative values we have presented an improved 
formulation of quantum reference frame transformation based on the 
use of the zero vector to `represent' the reference frame itself. In
addition, the transformations given for the qubit systems, as examples 
of transformations based on the relative notion of an observable with 
a discrete or finite spectrum, are new. So is the explicit application 
of the notion of noncommutative values, as well as its presentation, 
for a system with a finite dimensional Hilbert space. It is our hope 
that results for such systems may be more accessible experimentally. 
Together with the example of the the quantum, spatial transformation, 
we believe, they sketch a basic framework to write down any 
particular quantum reference frame transformations for any system. 

\section*{Appendix : Noncommutative Values of Position and Momentum 
Observables with the Schr\"odinger Wavefunction Representation}
With the position eigenstate basis $\hat{x} \left|x \rra= {x} \left|x \rra$,
we have $\left|\psi \rra =\int\!dx \left|x \rra\!\!\lla x|\psi \rra $ 
where the wavefunction $\psi(x)\equiv \lla x|\psi \rra$ is really 
an infinite set of coordinates for the physical state $\left|\psi \rra$
in the Hilbert space taken as a complex manifold. A (normalized)
position eigenstate with eigenvalue $x_o$  is given by the
wavefunction $\delta(x-x_o)$, and we have the matrix elements 
$(\hat{x})^{x'}_x = x \delta(x'-x)$.  From 
\bea
f_{\!\ssc\hat{x}} (\psi)= \frac{\int\!\!dx\; \bar\psi(x) x  \psi(x)}{ \int\!\!dx\; \bar\psi(x)  \psi(x)}
\eea
taken as a functional of the (normalized) wavefunction, we have 
the set of infinite coordinate derivatives can be expressed as the 
functional derivative 
\bea
V_{\!\ssc\hat{x}}(x)
 =\delta_\psi f_{\!\ssc\hat{x}}(\psi) \equiv \frac{\delta f_{\!\ssc\hat{x}}}{\delta \psi}(\psi)
= \bar\psi(x) (x -x_o) \;,
\eea
where $x_o$ here denotes the expectation value of
$f_{\!\ssc\hat{x}}$ evaluated for the fixed $\psi(x)$. Again,
we have really one value of $V_{\!\ssc\hat{x}}(x)$ at each $x$ 
value matching to the coordinate value of $\psi(x)$. For the 
momentum observable, we have
\bea
f_{\!\ssc\hat{p}} (\psi)
= \frac{\int\!\!dx\; \bar\psi(x) (-i\partial_x ) \psi(x)}{ \int\!\!dx\; \bar\psi(x)  \psi(x)}
= \frac{\int\!\!dx\; [i\partial_x \bar\psi(x)]  \psi(x)}{ \int\!\!dx\; \bar\psi(x)  \psi(x)}\;,
\eea
which gives 
\bea
V_{\!\ssc\hat{p}}(x)
 =\delta_\psi f_{\!\ssc\hat{p}}(\psi) 
=  (i \partial_x -p_o) \bar\psi(x)\;,
\eea
where $p_o$ again denotes the expectation value. In the same 
spirit, we read the matrix elements $(\hat{p})^{x'}_x$ from
\bea&&
\int\!\!dx'dx\; \bar\psi(x') (\hat{p})^{x'}_x \psi(x) \delta(x'-x)
= f_{\!\ssc\hat{p}} (\psi) 
= \int\!\!dx'dx\; [i\partial_{x'} \bar\psi(x')]  \delta(x'-x) \psi(x)
\sea\hspace*{.5in}
=  \int\!\!dx'dx\; \bar\psi(x')  [-i\partial_{x'}\delta(x'-x)] \psi(x) \;,
\eea
where $\partial_{x'} \delta(x'-x) = - \partial_{x} \delta(x'-x)$ are
derivatives of the delta function with respect to the variables,
and  we have the result $(\hat{p})^{x'}_x= -i\partial_{x'} \delta(x'-x) =i\partial_{x} \delta(x'-x)$.
We have then the matrix elements
\bea&&
(\hat{x}\hat{p})^{x'}_x=-i \int\!\!dy\; [\partial_{y}\delta(y-x)] y \delta(x'-y) = -i x'\partial_{x'} \delta(x'-x)\; 
\sea
(\hat{p}\hat{x})^{x'}_x= -i \int\!\!dy\; x \delta(y-x) \partial_{x'} \delta(x'-y)=-i x  \partial_{x'} \delta(x'-x)\;,
\sea
(\hat{x}\hat{p})^{x'}_x -(\hat{p}\hat{x})^{x'}_x =  -i (x'-x) \partial_{x'} \delta(x'-x) = i \delta(x'-x) \;,
\eea
and we also have
$f_{\!\ssc\hat{x}\hat{p}} = -i \int\!\!dx\; \bar\psi(x) x \partial_x \psi(x)$ and
$f_{\!\ssc\hat{p}\hat{x}} = -i \int\!\!dx\; \bar\psi(x) [x \partial_x \psi(x) + \psi(x) ]$.
The noncommutative product among the noncommutative
values corresponding to the formula of Eq.(\ref{ncx}) then gives
 \bea&&
V_{\!\ssc\hat{x}\hat{p}} (x)
 =   -i \int\!\!dy\; \bar\psi(y)  y \partial_y\delta(y-x) - \bar\psi(x) f_{\!\ssc\hat{x}\hat{p}} 
\sea\hspace*{.3in}
  = i \int\!\!dy\; \delta(y-x) [\bar\psi(y) + y \partial_y \bar\psi(y) ]- \bar\psi(x) f_{\!\ssc\hat{x}\hat{p}}
%\sea
  =  \bar\psi(x)  [i-f_{\!\ssc\hat{x}\hat{p}}]  + i  x \partial_x \bar\psi(x)\;,
\sea
V_{\!\ssc\hat{p}\hat{x}} (x)
 =  -i \int\!\!dy\; \bar\psi(y)  x \partial_y\delta(y-x) - \bar\psi(x) f_{\!\ssc\hat{p}\hat{x}}
\sea\hspace*{.3in}
  = i \int\!\!dy\; \delta(y-x)  x \partial_y \bar\psi(y) - \bar\psi(x) f_{\!\ssc\hat{p}\hat{x}}
%\sea
=  -\bar\psi(x)  f_{\!\ssc\hat{p}\hat{x}}  + i  x \partial_x \bar\psi(x)\;,
\sea
V_{\!\ssc\hat{x}\hat{p}} (x) - V_{\!\ssc\hat{p}\hat{x}} (x)
= \bar\psi(x)   [ i - f_{\!\ssc\hat{x}\hat{p}} + f_{\!\ssc\hat{p}\hat{x}}  ] =0
=\delta_\psi (f_{\!\ssc\hat{x}\hat{p}} - f_{\!\ssc\hat{p}\hat{x}}) 
= V_{\!\ssc\hat{x}\hat{p}-\hat{p}\hat{x}} (x) \;.
\eea
The last result is an important consistency check of the fact that the
noncommutative value of an observable as a product of two observables
is the product of the noncommutative values of the individual observables,
and noncommutative value of an observable as the commutator is the 
commutator of the individual noncommutative values.
We check further from
\bea&&
f_{\!\ssc\hat{x}\hat{p}} 
= f_{\!\ssc\hat{x}} f_{\!\ssc\hat{p}} + \int\!\!dx\; V_{\!\ssc\hat{x}}(x)  \bar{V}_{\!\ssc\hat{p}}(x) \;,
\sea
f_{\!\ssc\hat{p}\hat{x}} 
= f_{\!\ssc\hat{p}} f_{\!\ssc\hat{x}} + \int\!\!dx\; V_{\!\ssc\hat{p}}(x)  \bar{V}_{\!\ssc\hat{x}}(x) \;,
\eea
that
\bea&&
f_{\!\ssc\hat{x}\hat{p}}- f_{\!\ssc\hat{p}\hat{x}} 
= \int\!\!dx\;  \bigg[ V_{\!\ssc\hat{x}}(x)  \bar{V}_{\!\ssc\hat{p}}(x) - V_{\!\ssc\hat{p}}(x)  \bar{V}_{\!\ssc\hat{x}}(x) \bigg]
\sea
= \int\!\!dy\; \bigg[ \bar\psi(y) [y - f_{\!\ssc\hat{x}}] [-i \partial_y - f_{\!\ssc\hat{p}}]\psi(y) 
  - [i \partial_y - f_{\!\ssc\hat{p}}]\bar\psi(y) [y - f_{\!\ssc\hat{x}}] \psi(y) \bigg]
%\sea
%= \int\!\!dy\; \bar\psi(y) [y - f_{\!\ssc\hat{x}}] [-i \partial_y \psi(y) - f_{\!\ssc\hat{p}}\psi(y) ]
 % - [i \partial_y\bar\psi(y) - f_{\!\ssc\hat{p}} \bar\psi(y)] [y - f_{\!\ssc\hat{x}}] \psi(y) 
\sea
=  \int\!\!dy\; \bigg[  i f_{\!\ssc\hat{x}} [ \bar\psi(y)  \partial_y \psi(y) + \partial_y\bar\psi(y)   \psi(y) ]
 -i  [ \bar\psi(y) y \partial_y \psi(y) + \partial_y\bar\psi(y)  y \psi(y) ] \bigg]
\sea
= i \int\!\!dy\; \bigg[  \psi(y)  \partial_y [ \bar\psi(y) y ] - \partial_y\bar\psi(y)  y \psi(y) \bigg]=i \;.
\eea
Hence our noncommutative product expressions for the position and 
momentum observables are fully consistent. A generic observable would 
be taken as some functions of $\hat{x}$ and $\hat{p}$. The noncommutative 
expression hence gives definite results for at least all observables which can 
be expressed as a (noncommutative) polynomial in $\hat{x}$ and $\hat{p}$.  
Generalizing to the case of three pairs of $\hat{x}_i$ and $\hat{p}_i$ is straightforward.

Here, we give the details of the somewhat subtle calculation of 
section~\ref{sec3} for the momentum observable discussed at the end 
of the analysis for case (a). It shows results of the quantum spatial 
translation on the momentum observables not otherwise explicitly
illustrated. But is mostly to be read in reference to the above mentioned
analysis.
\bea &&
\hat{S}_x  \hat{p}_{\!\ssc B} \hat{S}_x^\dag
= \int\!\!dx''dy''dxdy\; \left|-x'',y''-x'' \rra_{\!\ssc AC}  \lla x'',y''\right|\hat{p}_{\!\ssc B} \left|x,y\rra_{\!\ssc BC}  \lla -x,y-x\right|_{\!\ssc AC} 
\sea
= \int\!\!dx''dy''dxdy\; [-i\partial_{x''}\delta(x''-x) ]\delta(y''-y)
\left|-x'',y''-x'' \rra\!\! \lla -x,y-x\right|_{\!\ssc AC} \;,
\sea
= \int\!\!dx'''dy'''dx'dy'\; [i\partial_{x'''}\delta(x'''-x')] \delta(y'''-y'-x'''+x')
\left|x''',y''' \rra\!\! \lla x',y'\right|_{\!\ssc AC} \;,
\eea
giving $\lla \phi' | \hat{S}_x  \hat{p}_{\!\ssc B} \hat{S}_x^\dag |\phi'\rra$ as
\bea &&%\hspace*{.3in}
%\lla \phi | \hat{S}_x  \hat{p}_{\!\ssc B} \hat{S}_x^\dag |\phi\rra = 
\int\!\!dx''dy''dxdy\; [-i\partial_{x''}\delta(x''-x)] \delta(y''-y) \delta(x''+x_o) \bar\psi(y''-x''+x_o) \delta(x+x_o) \psi(y-x+x_o)
\sea
= \int\!\!dx''dy''dxdy\; \delta(x''-x) \delta(y''-y) \delta(x+x_o) \psi(y-x+x_o) 
            i\partial_{x''} [\delta(x''+x_o) \bar\psi(y''-x''+x_o)]
\sea
= \int\!\!dxdy\;  \delta(x+x_o) [i\partial_{x}\delta(x+x_o)]|\psi(y-x+x_o)|^2
\sea\hspace*{1.2in}
   + \int\!\!dxdy\; \delta^2(x+x_o)   \psi(y-x+x_o) i  \partial_x\bar\psi(y-x+x_o)\;.
\eea
The first term is $-f_{\hat{p}_{\!\ssc A}}^f(\phi')$ and the second term is 
$-f_{\hat{p}_{\!\ssc C}}^f(\phi')$, which may be more easily seen through
substituting $y'=y-x$ and $x'=x$ into the direct expressions of 
\bea&&
f_{\hat{p}_{\!\ssc A}}^f(\phi') =  \int\!\!dx'dy'\; [-i\partial_{x'}\delta(x'+x_o)] \delta(x' +x_o)   |\psi(y'+x_o)|^2 
\sea
f_{\hat{p}_{\!\ssc C}}^f(\phi') =  \int\!\!dx'dy'\; \delta^2(x' +x_o) \psi(y'+x_o) i\partial_{y'}  \bar\psi(y'+x_o) \;.
\eea

%%%%%%%%%%%%%%%%%%%%%%%%%%%%%%%%
\section*{Acknowledgments :}%\setlength{\parindent}{0.0cm} 
The work is supported by the research grants number
109-2112-M-008-016 and 110-2112-M-008-016 of the MOST of Taiwan.

\end{document}